\documentclass[]{aa}  

\usepackage{natbib}
\usepackage{graphicx}
\usepackage{txfonts}
\usepackage{orcidlink}
\usepackage{multirow}
\usepackage{comment}

\makeatletter
\renewcommand*\aa@pageof{, page \thepage{} of \pageref*{LastPage}}
\usepackage{hyperref}
\hypersetup{colorlinks=true,allcolors=[rgb]{0,0,0.8}}
\hypersetup{
  colorlinks=true,
  allcolors=[rgb]{0,0,0.8},
  pdftitle={Stacked eROSITA Galaxy Groups},
  pdfauthor={Bulbul et al.},
}

\newcommand{\rosi}{eROSITA\xspace}

\newcommand{\erass}{eRASS1\xspace}

\newcommand{\ntot}{827 }

\usepackage{graphicx} % put this in your preamble

\newcommand{\apecdsOnepriTmean}{$0.96_{-0.02}^{+0.02}$}
\newcommand{\apecdsOnepriabundanc}{$0.11_{-0.01}^{+0.01}$}
\newcommand{\apecdsOneprinorm}{$9.50_{-0.40}^{+0.50}$}
\newcommand{\apecdsTwoprinorm}{$0.90_{-0.20}^{+0.10}$}
\newcommand{\gademdsOnepriTmean}{$1.03_{-0.04}^{+0.05}$}
\newcommand{\gademdsOnepriTsigma}{$0.37_{-0.10}^{+0.13}$}
\newcommand{\gademdsOnepriabundanc}{$0.18_{-0.03}^{+0.05}$}
\newcommand{\gademdsOneprinorm}{$8.40_{-0.60}^{+0.60}$}
\newcommand{\gademdsTwoprinorm}{$0.80_{-0.10}^{+0.10}$}
\newcommand{\gademGademdsOnepriTmean}{$0.96_{-0.04}^{+0.05}$}
\newcommand{\gademGademdsOnepriTsigma}{$0.28_{-0.10}^{+0.10}$}
\newcommand{\gademGademdsOneprinorm}{$6.30_{-1.40}^{+1.60}$}
\newcommand{\gademGademdsOnesecTmean}{$3.60_{-1.34}^{+1.58}$}
\newcommand{\gademGademdsOnesecTsigma}{$0.45_{-0.19}^{+0.34}$}
\newcommand{\gademGademdsOnepriabundanc}{$0.21_{-0.04}^{+0.06}$}
\newcommand{\gademGademdsOnesecnorm}{$1.60_{-1.20}^{+0.80}$}
\newcommand{\gademGademdsTwoprinorm}{$0.44_{-0.30}^{+0.30}$}
\newcommand{\gademGademdsTwosecnorm}{$0.38_{-0.30}^{+0.40}$}
\newcommand{\gademBremsdsOnepriTmean}{$0.98_{-0.04}^{+0.05}$}
\newcommand{\gademBremsdsOnepriTsigma}{$0.31_{-0.10}^{+0.12}$}
\newcommand{\gademBremsdsOnepriabundanc}{$0.24_{-0.06}^{+0.10}$}
\newcommand{\gademBremsdsOneprinorm}{$6.20_{-1.81}^{+1.50}$}
\newcommand{\gademBremsdsOnebremsTmean}{$4.21_{-2.33}^{+7.40}$}
\newcommand{\gademBremsdsOnebremsnorm}{$0.50_{-0.32}^{+0.40}$}
\newcommand{\gademBremsdsTwoprinorm}{$0.40_{-0.21}^{+0.30}$}
\newcommand{\gademBremsdsTwobremsnorm}{$0.12_{-0.09}^{+0.10}$}
\newcommand{\gademPLdsOnepriTmean}{$0.99_{-0.04}^{+0.05}$}
\newcommand{\gademPLdsOnepriTsigma}{$0.34_{-0.10}^{+0.11}$}
\newcommand{\gademPLdsOnepriabundanc}{$0.22_{-0.05}^{+0.09}$}
\newcommand{\gademPLdsOneprinorm}{$6.88_{-1.88}^{+1.20}$}
\newcommand{\gademPLdsTwoprinorm}{$0.46_{-0.17}^{+0.27}$}
\newcommand{\gademPLdsTwoplPhoIndex}{$1.41_{-0.69}^{+0.40}$}
\newcommand{\gademPLdsOneplPhoIndex}{$1.41_{-0.69}^{+0.40}$}
\newcommand{\gademPLdsOneplnorm}{$0.23_{-0.17}^{+0.27}$}
\newcommand{\gademPLdsTwoplnorm}{$0.07_{-0.04}^{+0.07}$}

\begin{document}

%\title{The SRG/eROSITA All-Sky Survey}
\title{Thermal or Non-thermal? Diffuse emission in the infall region of stacked galaxy groups }

\author{
E. Bulbul\inst{1,2}\orcidlink{0000-0002-7619-5399}, 
X. Zhang\inst{1},
Z. Ding\inst{1}, 
T. Mistele\inst{1}\orcidlink{0000-0001-7048-3173},
M. Kluge\inst{1}, 
E. Artis\inst{1}\orcidlink{0009-0001-6055-8503},
Y. E. Bahar\inst{3},
K. Dennerl\inst{1}, 
D. Eckert\inst{4},
L. Fiorino\inst{1}\orcidlink{0009-0008-8885-6909},
P. F. Hopkins\inst{5},
N. Malavasi\inst{1}\orcidlink{0000-0001-9033-7958},
A. Merloni\inst{1},
K. Nandra \inst{1}, 
E. Quataert \inst{6},
M. E. Ramos-Ceja\inst{1}\orcidlink{0000-0002-9117-3251}, 
J. S. Sanders\inst{1}, 
J. Strunk\inst{1}, \and
S. Zelmer\inst{1}
}

\institute{
    Max Planck Institute for Extraterrestrial Physics, Giessenbachstrasse 1, 85748 Garching, Germany\label{inst:mpe}\\
          \email{ebulbul@mpe.mpg.de}
    \and 
    Faculty of Physics, Fakult\"at f\"ur Physik, Ludwig-Maximilians-Universit\"at, Scheinerstr. 1, 81679 M\"unchen, Germany
     \and
    INAF, Osservatorio di Astrofisica e Scienza dello Spazio, via Piero Gobetti 93/3, I-40129 Bologna, Italy  
    \and
    Department of Astronomy, University of Geneva, Ch. d’Ecogia 16, CH-1290 Versoix, Switzerland
    \and
    Division of Physics, Mathematics, and Astronomy, California Institute of Technology, Pasadena, CA 91125, USA
    \and
    Department of Astrophysical Sciences, Princeton University, Princeton, NJ 08544, USA
    }

\date{Received ---; accepted ---}

\titlerunning{X-ray Emission in the Infall Regions of Galaxy Groups}
\authorrunning{Bulbul et al.}

\abstract{
The faint infall regions surrounding the virial radius of galaxy groups remain largely unexplored due to their low X-ray surface brightness. Using the large statistical power of SRG/eROSITA survey observations, we present the first spectroscopic measurement of the intragroup medium (IGrM) in the infall regions of a large sample of low-mass galaxy groups ($M_{\rm tot}<1\times10^{14}\,M_{sun}$), extending to $\sim2\,R_{200m}$ (2.2~Mpc). Through spectral stacking of \ntot nearby groups from the first eROSITA All-Sky Survey catalog, we detect diffuse emission and measure the thermodynamic properties of gas at densities previously inaccessible to X-ray observations. The stacked spectra are well described by a Gaussian differential emission measure model, yielding a temperature distribution with a mean temperature of \gademGademdsOnepriTmean~keV and width of \gademGademdsOnepriTsigma~keV, and a metal abundance of \gademGademdsOnepriabundanc~A$_{sun}$, consistent with expectations for group outskirts. The inferred electron densities decrease from $(4.8\pm1.3)\times10^{-5}$~cm$^{-3}$ at $(0.7-2)\,R_{500c}$ to $(5.5\pm2.0)\times10^{-6}$~cm$^{-3}$ at $(2-4)\,R_{500c}$, demonstrating eROSITA’s ability to probe the low-density outskirts of galaxy groups. Residual emission in the spectra suggests the presence of an additional spectral component. While a secondary thermal interpretation requires an unexpectedly hot, metal-poor plasma, a non-thermal inverse Compton model provides an equally plausible explanation, contributing $\sim30\%$ of the thermal flux. 
Assuming that the additional component is produced by inverse Compton emission from a common population of relativistic electrons, the inferred magnetic field strength would be in the sub-$\mu$G regime. These results demonstrate the potential of eROSITA to reveal faint thermal and non-thermal processes associated with the formation and evolution of large-scale structure.
}


\maketitle
\section{Introduction}

The large-scale structure and the cosmic web are composed of smaller masses of galaxies, galaxy groups, and massive galaxy clusters, embedded in filaments. Originating from primordial density perturbations, baryonic matter evolves gravitationally to form galaxy groups, serving as a link between small-scale galaxies and large-scale massive clusters. Bridging the mass scale between individual galaxies and massive galaxy clusters, galaxy groups occupy a critical regime for understanding structure formation. Owing to their abundance in recent surveys, such as eROSITA aboard the Spectrum Roentgen Gamma (SRG) mission \citep{Sunyaev2021, Predehl2021}, they offer substantial statistical power for investigating astrophysical processes within gravitationally collapsed halos \citep{Bulbul2024, Bahar2024}. Their relatively shallow gravitational potentials facilitate the redistribution of baryons within and beyond the host halo, making them uniquely sensitive probes for exploring non-gravitational processes.

In galaxy groups, the intragroup medium (IGrM) is composed of hot (kT$\,\sim 10^{7}$~K) gas that permeates the space between galaxies and emits predominantly via thermal bremsstrahlung in the X-ray band. Over time, this radiative cooling of the plasma is expected to lead to the growth of supermassive black holes and significant star formation in central galaxies \citep{McNamara2007, Fabian2012}. A heating mechanism that can provide sufficient heating of the ambient halo gas to counterbalance cooling is active galactic nucleus (AGN) feedback, in which the energy released by the central supermassive black hole reheats the surrounding gas, suppressing further cooling \citep{Randall2018, Sijacki2007}. Strong AGN feedback can expel baryons around and beyond the virial radius, redistributing gas into the outer regions of galaxy groups and clusters even out to their Virial radii \citep{Springel2018, Schaye2023, Bigwood2025}. Recent kSZ and X-ray observations provide evidence for significant gas displacement around the Virial radius in low-mass groups, highlighting the well-extended nature of the IGrM medium \citep{Hadzhiyska2024, Siegel2025, Popesso2024}. However, the underlying physical nature of this extended emission remains elusive, as current observations are unable to distinguish between a displaced thermal gas component and non-thermal contributions from cosmic-ray populations accelerated by merger or accretion shocks.

While the redistribution of thermal plasma via feedback-driven outflows or heated gas through accretion shocks provides a plausible explanation, a non-thermal origin represents a compelling alternative. In this scenario, the signal around the Virial radius may be dominated by non-thermal processes rather than high-temperature thermal plasma displaced by the central AGN. In hierarchical structure formation, accretion shocks arise near the virial radius as IGrM material accretes into the gravitational potential. These shocks accelerate electrons via diffusive shock acceleration, producing non-thermal populations that upscatter CMB photons through inverse Compton processes. The low-energy electrons (0.1–1~GeV) with much longer lifetimes ($\sim\, 10^{9}$ yr) can stream out to 100~kpc, persisting long after an AGN has turned off. By up-scattering CMB photons into the X-ray band via Inverse Compton scattering, these electrons may create diffuse X-ray plasma at halo outskirts. The resulting emission can extend over large spatial scales and may appear from optical to X-ray and $\gamma$-ray energies, depending on the electron population. Recent theoretical studies suggest that Inverse Compton (IC) emission from cosmic-ray electrons may contribute non-negligibly to the observed X-ray surface brightness in both cores and outskirts of galaxies, and galaxy groups \citep{Quataert2025, Hopkins2026}. Many observational campaigns using X-ray satellites have therefore primarily yielded upper limits on IC fluxes \citep{Wik2012, Wik2014}.

The baryon distribution is well studied in the cores of several prominent galaxy groups through multi-wavelength analyses \citep[e.g.,][]{Sun2009, OSullivan2011b, Randall2015, Bahar2024, Eckert2025, Mernier2023}. However, observational constraints in infall regions of galaxy groups beyond $>\mathrm{R}_{500c}$ remain scarce. This is primarily due to the low temperature and low electron densities ($\sim\,10^{-4}$ to $10^{-6}$ cm$^{-3}$) in these regions, which significantly reduce the X-ray surface brightness, making the direct signatures of non-gravitational feedback exceptionally challenging to measure. The X-ray instruments, covering a wide band pass, enable the measurement of the harder, non-thermal or thermal X-ray emission arising from the inverse Compton scattering of electrons accelerated by galaxy group mergers, AGN, or acceleration shocks. The thermal emission from the intra-group medium peaks in the soft X-ray band below 2~keV, allowing non-thermal components, if present, to be constrained in the hard X-ray band above this energy. 

Although X-ray instruments have been used to study a few selected nearby bright galaxy groups, a large statistical sample has only become available through the launch of eROSITA. In its first All-Sky Survey, eROSITA has detected 12,247 galaxy clusters and groups in the Western Galactic Hemisphere \citep {Bulbul2024, Kluge2024, Merloni2024}. Included among these is a well-defined and uniformly selected subsample of 1178 galaxy groups drawn from the parent sample. This uniformly galaxy group sample enables the analysis of X-ray–derived entropy profiles of the IGrM using population modeling techniques, offering substantial statistical power to test and constrain AGN feedback models \citep{Bahar2024}. Beyond providing a large, uniformly selected sample, this dataset provides sufficient statistical leverage through stacking analyses to constrain weak X-ray emission in the infall region of galaxy groups that would otherwise fall below the detection limits of targeted X-ray observations \citep[see e.g.,][]{Zhang2024}. Additionally, the wide energy band of eROSITA is well-suited for detecting X-ray emission from both thermal and non-thermal plasma components. As such, the continuum and line emission from galaxy groups typically peaks below 2~keV, where eROSITA has its highest sensitivity. In addition, the coverage of the hard X-ray band enables constraints on potential power-law emission in the hard band $>2$~keV, such as that arising from inverse Compton scattering.

In this study, we use a sample of \ntot nearby eROSITA-selected galaxy groups to perform a comprehensive X-ray stacking analysis, incorporating spectral data from the deepest eROSITA All-Sky Survey observations. The sample selection is highly robust, achieving a purity of 98\% \citep{Bahar2024}. Data were gathered from 4.4 consecutive stacked survey passes conducted between 2019 and 2022, hereafter referred to as eRASS:5. Our analysis focuses on the spectroscopic properties of the plasma in the group infall region. Specifically, we stack spectra around the Virial radius out to 2$R_{200m}$\footnote{R$_{200m}$ is the overdensity radius within which the local density of the intracluster medium (ICM) is 200 times the mean density of the Universe at the cluster redshift.} to investigate plasma properties at the transition between the virialized intra-group medium and the large-scale diffuse filaments. This paper is organized as follows: In Section~\ref{sec:analysis}, we describe the selection of the group sample, X-ray analysis, and stacking technique. We present our results and their interpretation together with systematics in Sections~\ref{sec:results} and \ref{sec:discussion}, and provide our conclusions in Section~\ref{sec:conclusions}. The quoted statistical uncertainties correspond to a 1-$\sigma$ confidence level unless noted otherwise. We adopt a flat $\Lambda$CDM cosmology with parameters
H$_{0} = 67.3$~km~s$^{-1}$~Mpc$^{-1}$, $\Omega_M =  0.315$, and $\Omega_{DE} = 0.685$ throughout the analysis \citep{Planck2020}.

\section{Sample and Data Analysis}
\label{sec:analysis}
In this work, we utilize a robust sample of galaxy groups initially detected during the first eROSITA All-Sky Survey (eRASS1) to perform a high-signal-to-noise stacking analysis. By leveraging the full depth of the eRASS:5 data set, comprising five consecutive survey passes combined, we extract spectra and derive detailed temperature information across the sample. The following sections outline our methodology, including the criteria for sample selection, the X-ray processing, and the specific implementation of the spectral and spatial stacking procedures.
\begin{figure}
\begin{tabular}{c}
\includegraphics[width=0.5\textwidth]{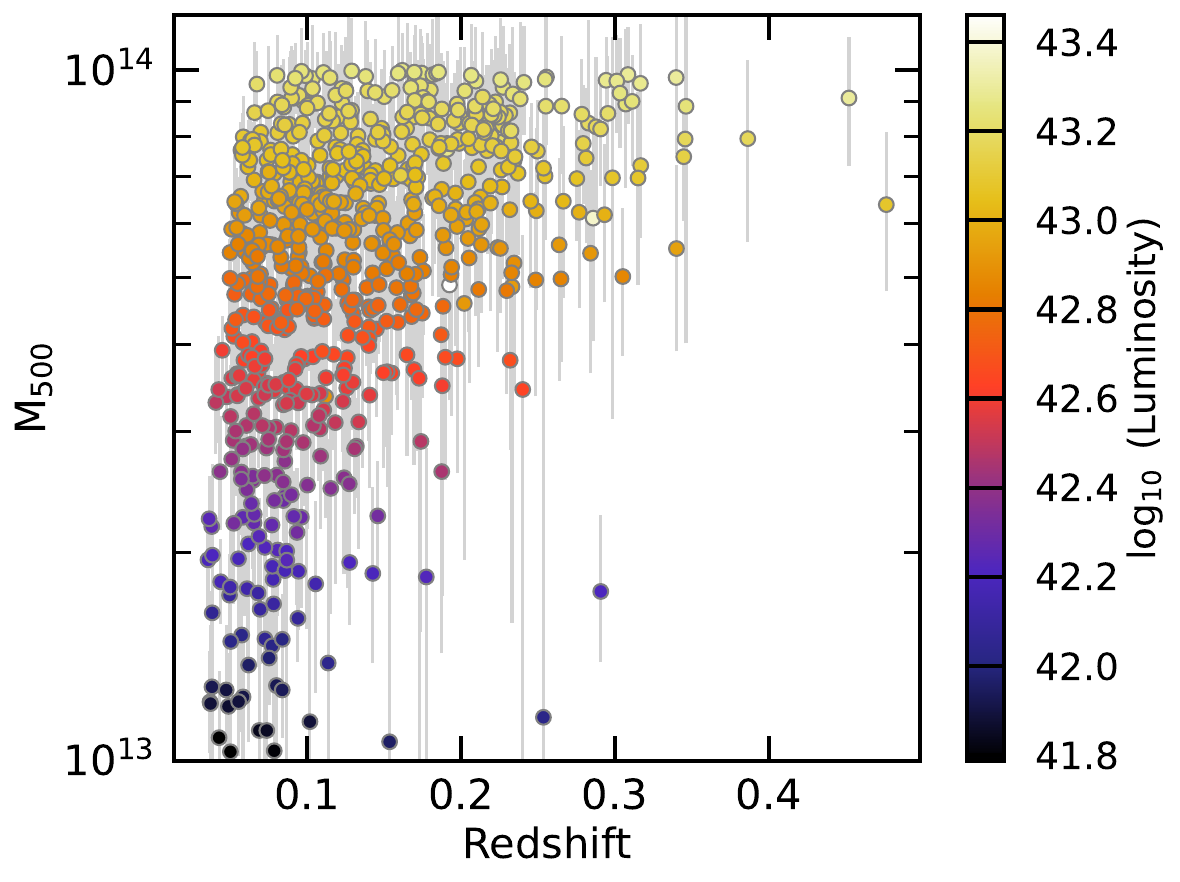} 
\end{tabular}
\caption{Mass and redshift distributions of the \ntot\ galaxy groups detected in the first eROSITA All-Sky Survey (Western Galactic Hemisphere) and included in this study. The sample spans a redshift range of 0.003–0.5, with a median mass of $5.91\times10^{13}$~M$_{\odot}$ and a median redshift of 0.15.\label{fig:sample}}
\end{figure}
\subsection{Sample Selection}
\label{sec:sample}

\begin{figure}
\begin{tabular}{c}
\includegraphics[width=0.45\textwidth]{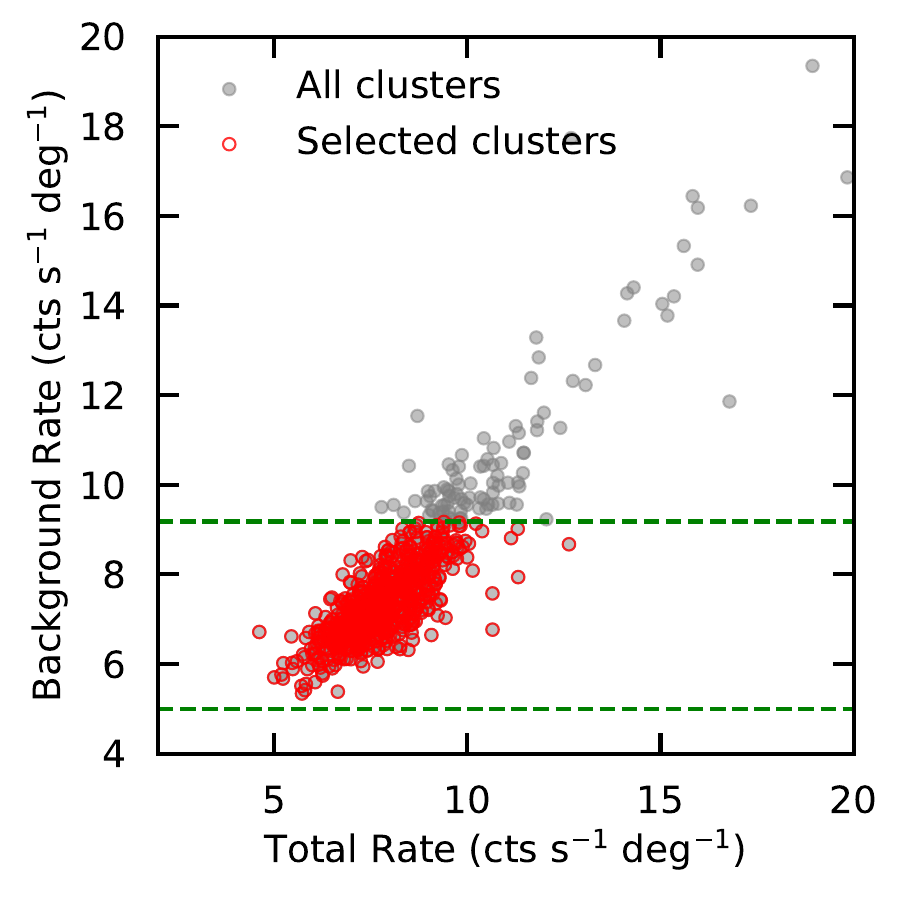} 
\end{tabular}
\caption{Total source count rate measured within (0.7–2)~R$_{500c}$ is shown as a function of the background count rate estimated in the (4–6)~R$_{500c}$ annulus for the parent galaxy group sample \citep{Bulbul2024, Bahar2024}. Gray points denote the full sample, while red points highlight systems that satisfy the background-filtering criteria and are retained for stacking. The horizontal green dashed lines indicate the adopted background selection thresholds. This filtering based on background count rate yields a final sample of \ntot\ galaxy groups. \label{fig:filtering}}
\end{figure}

The galaxy group sample is originally derived from the primary cluster and group catalog of the first eROSITA All-Sky Survey (eRASS1) within the Western Galactic Hemisphere, using an initial detection criterion of extent likelihood of $\mathcal{L}_{\text{ext}}>3$ \citep{Bulbul2024, Kluge2024}. While the eRASS1 primary catalog contains 2,950 objects with total masses (M$_{500}$\footnote{The total mass enclosed within the overdensity radius R$_{500c}$}) below $10^{14}$~M$_{\odot}$, the low-richness ($\lambda<20$) and low-extent likelihood regimes exhibit contamination levels exceeding 15\% \citep{Kluge2024}. To mitigate the removal of spurious sources and AGN, \citet{Bahar2024} implements a more stringent likelihood threshold of $\mathcal{L}_{\text{ext}} > 5.5$ and performs visual inspections using deeper eRASS:5 data. Given that genuine galaxy groups at the sample's median redshift (z$\sim$0.11) are expected to be spatially extended, 841 compact sources with EXT$<$20 arcsec are removed. As a final step, the visual inspection is performed, yielding a refined sample of 1,078 groups, shown in Figure~\ref{fig:sample}. To avoid a potential bias, we exclude sources located near survey edges and those with unusually high or low background rates, specifically those outside the 10th-90th percentile range. The total, background, and filtering rates applied to the sample are shown in Figure~\ref{fig:filtering}. Our final sample consists of \ntot galaxy groups, spanning a redshift range of $0.003<z<0.5$, with a median mass of $5.91\times10^{13}$ M$_{\odot}$ and a median redshift of 0.10. The distributions of mass, redshift, and X-ray luminosity for this sample are illustrated in Figure~\ref{fig:sample}. The overdensity radius of R$_{200m}$ corresponds to 2.3~R$_{500c}$ at the median redshift of this sample. The median R$_{500c}$ for the sample is 567.0~kpc.

\subsection{X-ray Analysis and Spectral Extraction}
We apply the standard data reduction and analysis procedure to eROSITA data using standard analysis methods \citep{Merloni2024, Bulbul2024, Liu2022}. The stacked event files from eRASS:5 collected between 12th December 2019 and 19th December 2021 are processed using the eROSITA Science Analysis Software System \citep[eSASS;][]{Brunner2022}, with pipeline version 020 \citep{Merloni2024}. The events from TM8 (including Telescope Modules 1, 2, 3, 4, and 6) are used in this analysis to avoid soft-band contamination caused by the optical light leak, which would produce emission peaks in galaxy groups. The data analysis has utilized the software and calibration packages of the eSASS version \texttt{eSASSusers\_211214\_0\_4} to generate eROSITA event files, images, exposure maps, and spectra \citep{Predehl2021}. Source and background spectra and response files are generated using the {\tt srctool} command. The exposure keywords generated by the {\tt srctool} command correspond to the source's average livetime. To compute the accurate on-source exposure, these keywords are adjusted using the correction factor provided by the "CORRPSF" keyword in the ARF files. Images and exposure maps of TM8 (all detectors combined) for each cluster are generated using the {\tt evtool} and {\tt expmap} tools. All images are visually inspected to ensure that both extended and point sources are excluded from subsequent analysis. Although the eROSITA detection algorithm ({\tt ermldet}) is self-consistent and the eRASS:5 point source catalogs are employed for cleaning, some extended sources with X-ray emission may be missed by the detection procedure \citep[see][]{Brunner2022, Liu2022}. Such cases constitute less than 1\% of the sample. Visual inspection is therefore performed to identify and remove these residual sources before further analysis.

\subsection{Stacking Method for X-ray Spectra}
\label{sec:stacking}

To measure spectral properties such as the plasma temperature, we stack the \rosi\ spectra of galaxy groups in the \erass\ sample. For the stacking analysis, we follow the method developed by \citet{Bulbul2014}, which was first applied to \rosi\ data by \cite{Zhang2024}. Stacking a large number of clusters spanning a range of redshifts requires shifting them to a common redshift. Therefore, each spectrum is first shifted to the source's rest frame to enhance the source signal. The ARFs and RMFs are remapped to the rest frame of the source. An additional advantage of this approach is that background features, including fluorescent lines and absorption edges, become smeared out in the final stacked spectrum and can thus be more easily removed from the total spectrum, as they would also be prominent in the stacked background spectra.

The effective area curves in the ARF files output by srctool were multiplied by a correction curve as a function of energy to account for file calibration updates. These updates consist of the updated vignetting function and on-axis effective area, which will be included in the next major processing of the eROSITA data (version 040). The new vignetting function was computed by stacking sources as a function of instrument off-axis angle and energy (Sanders et al., in prep.). The correction curve assumes that sources are distributed uniformly over the field of view, which is a good approximation for the eROSITA survey data. The on-axis ARF model was updated to better fit sources using a standardized International Astronomical Consortium for High Energy Calibration (IACHEC) model and to be compatible with ground data measured using the MPE Panter facility \citep{Dennerl2021}. 

We use the same event files to extract images. The cleaned images are carefully inspected to ensure that all relevant extended and point sources are properly removed during spectral extraction. The event files are then blueshifted to the rest frame by adjusting the PI column using the redshifts reported in \citet{Kluge2024}. Using the eSASS tool \texttt{srctool}, spectra are extracted from the radial regions $0.7$–$2~R_{500\mathrm{c}}$, $2$–$4~R_{500\mathrm{c}}$, and $4-6~R_{500\mathrm{c}}$. The R$_{500c}$ values are adopted from the \erass\ primary cluster catalog \citep{Bulbul2024, Kluge2024} and are derived from weak-lensing–calibrated mass measurements assuming an \rosi\ cosmology \citep[see][for details of the mass calibration]{Ghirardini2024, Grandis2024, Kleinebreil2025, Okabe2025, Chiu2025}.

The response files generated with \texttt{srctool} are stacked using the \texttt{HEASOFT} tools \texttt{addarf} and \texttt{addrmf}. Each ancillary response file is corrected for the corresponding Galactic column density using the HI4PI survey in a self-consistent manner, with the Galactic neutral hydrogen column density computed to reduce uncertainties from absorption in the stacked spectral fitting \citep{HI4PI2016}. The stacking procedure employs a weighting scheme based on the relative source counts to enhance the signal-to-noise ratio of the IGrM emission. To model the background components, including the soft X-ray background, cosmic-ray–induced background, and detector background, spectra are extracted from an annular region spanning $4-6$~R$_{500c}$ centered on the group X-ray peak. In addition, the point sources are removed prior to spectral extraction to prevent contamination of the diffuse emission. The total and background spectra are subsequently stacked in the rest frame using the \texttt{HEASOFT} tool \texttt{mathpha}. After background subtraction, the resulting stacked spectra contain approximately $1.3\times10^{6}$, $5.9\times10^{6}$, and $4.4\times10^{6}$, $7.4\times10^{6}$ total counts in stacked exposure times of 3.5~Ms, 5.1~Ms, 7.2M~s and in the radial ranges of $0.7$–$2~R_{500\mathrm{c}}$, $2$–$4~R_{500\mathrm{c}}$, and $4-6~R_{500\mathrm{c}}$ in the 0.3-7~keV band.

\subsection{Temperature Distribution of the Sample}
\label{sec:tempdist}

Understanding the sample's properties is essential for interpreting the results in this paper. Before we present the fitting result, we first study in depth the expected plasma temperature of the sample. Due to the faint nature of spectra in the infall regions of low-mass groups, spectra of individual faint galaxy groups lack sufficient statistics to constrain plasma properties via X-ray spectroscopy with current X-ray telescopes, necessitating stacking analyses in these regions \citep[see][for massive clusters, and cosmic filaments]{Zhang2024, Zhang2025}. To aid in the interpretation of the results presented later, we examine the expected temperature distribution of the galaxy groups included in the stacked sample.

\begin{figure}
\begin{tabular}{c}
\includegraphics[width=0.48\textwidth]{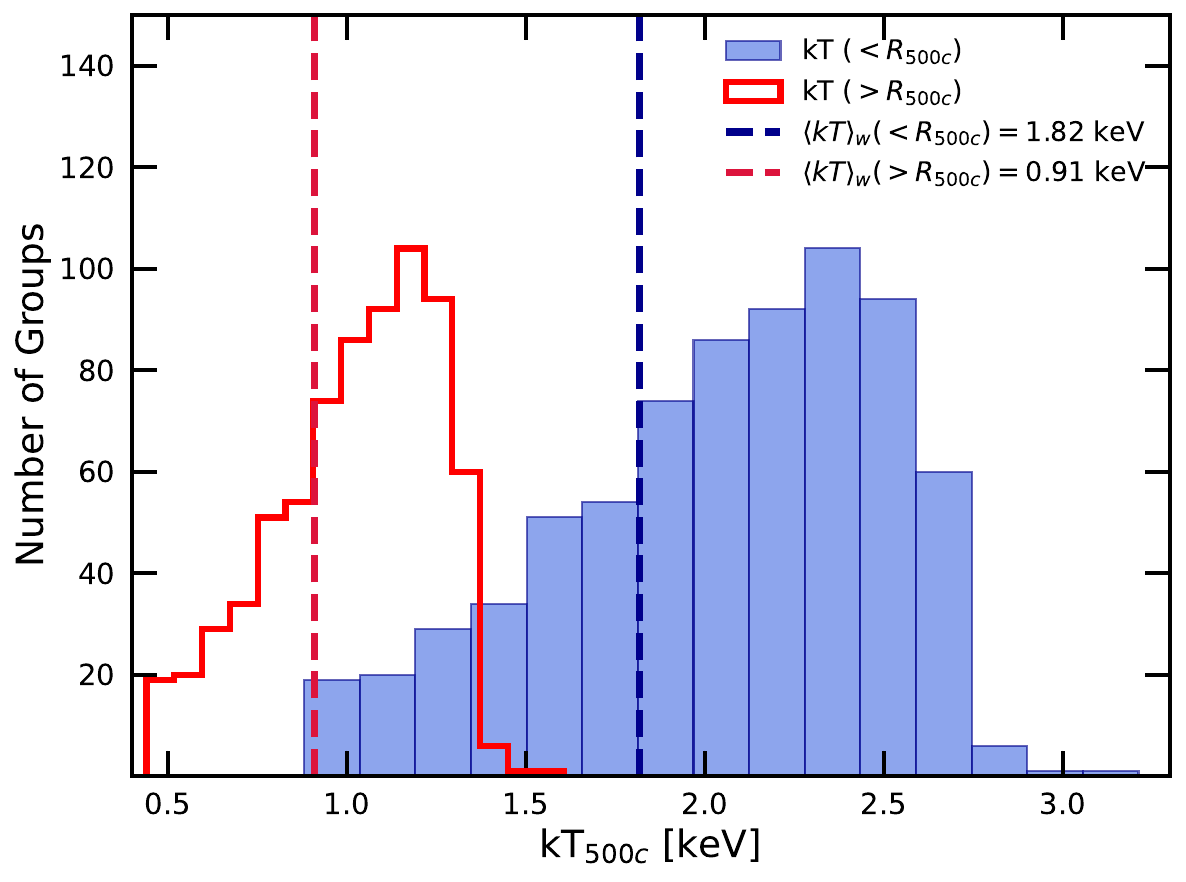} 
\end{tabular}
\caption{The temperature distribution of the sample within R$_{500c}$ is shown in blue. The expected mixed temperature distribution, comprising equal contributions from within and beyond R$_{500c}$, is represented by the red histogram. The vertical dashed lines indicate the weighted temperatures, calculated using the same weighting scheme applied in the stacking analysis. Overall, in the outer regions, no plasma component with temperatures exceeding $1.5~\mathrm{keV}$ is anticipated.\label{fig:sample_kT}}
\end{figure}

Because individual temperature measurements cannot be reliably obtained from the low signal-to-noise data, we instead use the luminosities of the groups reported in \citet{Bulbul2024} and apply the X-ray luminosity temperature ($L_{X}$–$kT$) scaling relations derived from the eROSITA selection-based samples \citep{Bahar2022, Liu2022, Bulbul2022, Ramos-Ceja2025}. Given that the eFEDS sample is among the most suitable eROSITA-selected samples with a larger fraction of galaxy groups with a similar selection, these scaling relations are well-suited to represent the properties of our dataset. The inferred temperature distribution within R$_{500c}$ is shown in blue in Figure~\ref{fig:sample_kT}, with a weighted mean temperature of 1.8~keV. Here, to calculate the weighted mean, we use the same weighting scheme as in the stacking analysis.

At larger radii, at R$_{500c}$, temperatures are reported to be approximately half the peak values in previous studies \citep{Rasia2006, Sun2009, Mernier2023}. The distribution shown in red in Figure~\ref{fig:sample_kT} represents the expected temperature distribution around and beyond R$_{500c}$, with a weighted mean temperature of 0.9~keV, indicated in dashed lines. These estimates serve as reference values for understanding the sample, evaluating, and interpreting results from the spectral fitting analysis.

\section{Results}
\label{sec:results}
In this section, we present the spectroscopic characterization of the plasma detected in the infall regions of galaxy groups around and beyond R$_{500c}$. We report the results of the spectral modeling and discuss their implications for the physical properties of the plasma in the infall regions of galaxy groups beyond 
$R_{500c}$.

The stacked source spectra extracted from the radial ranges $0.7$--$2~R_{500\mathrm{c}}$ and $2$--$4~R_{500\mathrm{c}}$, after background subtraction, together with the best-fitting models and residuals, are shown in Figure~\ref{fig:spec_1t}. 
We refer to these regions as Regions 1 and 2, hereafter. For visualization purposes, the data are rebinned to achieve a minimum signal-to-noise ratio of 15 per bin. The total background component extracted from 4--6R$_{500c}$ is subtracted prior to fitting. The local background component includes the non–X-ray background (NXB), the soft thermal foreground emission associated with the Milky Way, and the contribution from unresolved AGN. The Milky Way foreground component is completely subtracted successfully, with no leftover charge emission from O~VII and O~VIII ions (at $\sim$0.56~keV and $\sim$0.65~keV) observed in the net source spectrum \citep{Kuntz2015}. These lines are mostly prominent in the local foreground. Similarly, the detector background is removed from the net spectrum; the Fe-K line at 6.4~keV is fully removed, and no significant emission from non-X-ray background features remains \citep{Freyberg2021, Predehl2021}. The cosmic X-ray background emission from unresolved sources is also subtracted, as the background region surrounding the cluster includes this component. As a final check, we verify that the hard-band count rates in the 7-10~keV bandpass, where the spectra are dominated by the Cosmic X-ray Background (CXB), instrumental, and detector backgrounds, are consistent between the source and background regions. In all cases, this condition is satisfied, indicating that these background components have been successfully subtracted. After background subtraction, the stacked spectra yield 42,000 counts, with 15,400 net source counts in Regions 1 and 2, respectively.

Although our stacked spectra are not limited by photon statistics, we adopt the C-statistic in XSPEC and perform the spectral fitting using the Bayesian X-ray Analysis (BXA) framework \citep{Buchner2014, Kaastra2017}. We first fit the spectra extracted from Regions~1 and 2 with thermal models. The fitting is restricted to the $0.3$–$6.0$~keV energy range. For the modeling, the spectral fit to the background-subtracted spectra is performed using the thermal model(s) \texttt{APEC} \citep{Foster2012}, assuming an abundance table of \cite{Lodders2003}. 
In this work, we focus exclusively on the spectroscopic properties of the plasma in regions at and beyond R$_{500c}$, with particular emphasis on the infall regions of galaxy groups. A more comprehensive thermodynamic characterization of these regions will be presented in a forthcoming study (Bulbul et al. 2026, in preparation).

\subsection{Modeling the Individual regions around and beyond R$_{500c}$}
\label{sec:individualFits}

Following background subtraction of the total spectrum (and absorption correction to the ancillary response files), the source spectra are modeled using a single-temperature \texttt{APEC} model, applied independently to two radial ranges: 0.7--2~R$_{500c}$ (Region~1) and 2--4~R$_{500c}$ (Region~2). In each region, the temperature, metallicity, and normalization are allowed to vary freely, while the redshift is fixed to zero to represent the rest-frame emission of the stacked spectra. Allowing redshift to vary does not significantly change the goodness of the fit; the best-fit redshift remains consistent with zero, indicating that the blue-shifting and stacking method worked as expected. 

\begin{figure}
\begin{tabular}{c}
\includegraphics[width=0.48\textwidth]{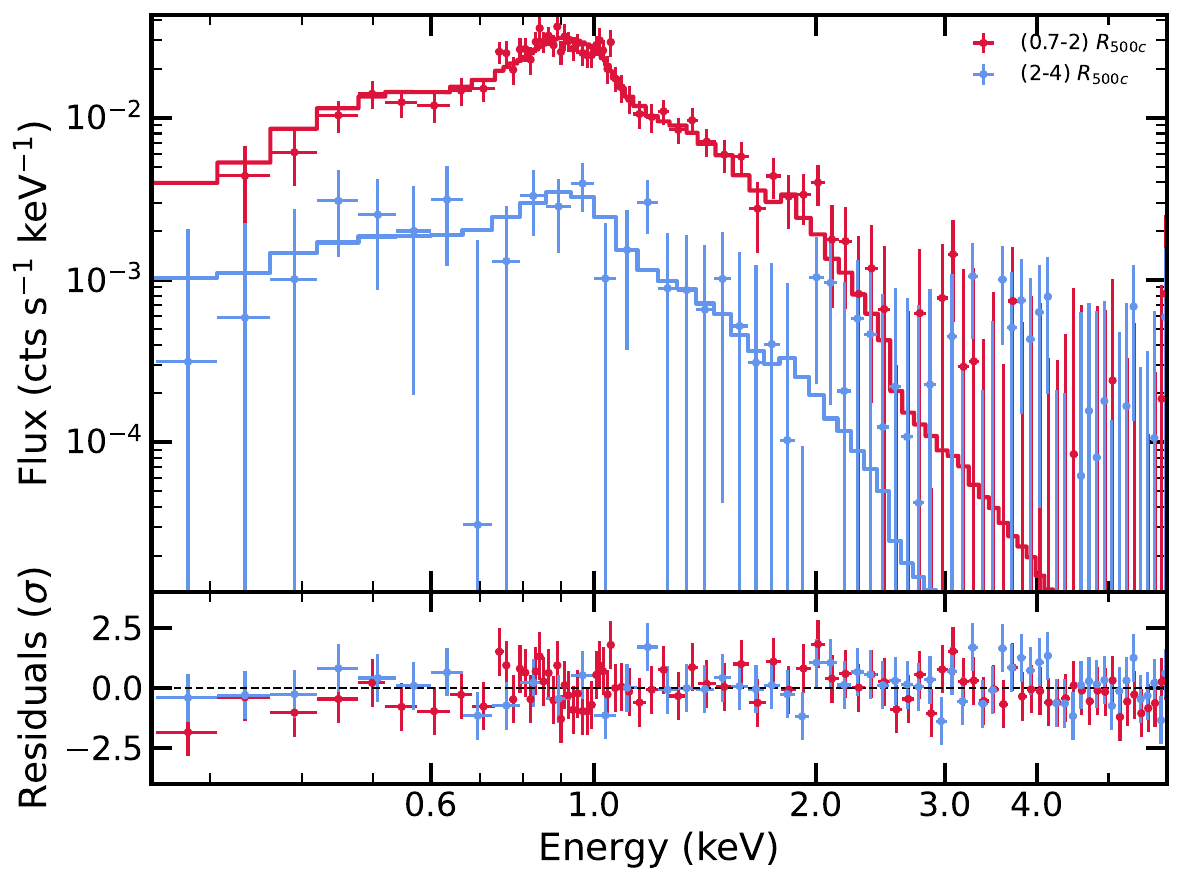 } 
\end{tabular}
\caption{The stacked X-ray spectrum, extracted from the deepest eROSITA observations, is shown for the radial ranges (0.7–2)~R$_{500c}$ (red) and (2–4)~R$_{500c}$ (dark blue) for the combined sample of \ntot galaxy groups, after subtraction of both the astrophysical X-ray and instrumental background components. The best-fit single-temperature thermal models (dashed lines) are overplotted in the bottom panel. Residuals observed in the soft and hard X-ray bands in both radial bins suggest the presence of an additional plasma component.\label{fig:spec_1t}}
\end{figure}

In the spectra extracted from Region~1, (0.7-2)~R$_{500c}$, the best-fit model yields a temperature of \apecdsOnepriTmean~keV and a metal abundance of $0.12\pm0.01$~$A_{sun}$, and a normalization of $5.76 \pm 0.31\times 10^{-5}$~cm$^{-5}$. 
The derived parameters are broadly consistent with expectations for these regions; the overall fit quality is good, with a C-statistic of 500 with 601 degrees of freedom (d.o.f.). We observe slight residuals in the spectra, suggesting the presence of an additional spectral component, shown in Figure~\ref{fig:spec_1t}.

Next, we fit the outermost Region~2 ($2-4$~R$_{500c}$), which approximately corresponds to the Virial radius, 0.8-1.7~R$_{200m}$, for this sample at the median redshift. Thanks to eROSITA’s soft X-ray sensitivity and large collecting area, we are able, for the first time, to characterize emission around R$_{200m}$ in galaxy groups using stacked eROSITA spectra, a regime that has not previously been accessible \citep[see also][ for the cluster regime]{Zhang2025}. A single-temperature model yields a temperature of $0.97 \pm 0.18$~keV and a metal abundance is not constrained, the 68\% upper limit on the measurements is $<0.13$~A$_{\odot}$, with a normalization of $(1.11 \pm 0.32)\times 10^{-5}$~cm$^{-5}$.
Similarly, in the inner region, Region~1, the residuals below 1.5~keV, particularly around the Fe-L complex, along with a C-statistic value of 571.9 for 601 degrees of freedom, may indicate the presence of an additional spectral component. 

The departures from the model in the residuals of the data shown in the left panel of Figure~\ref {fig:spec_1t}, where the 1-temperature fits to the stacked data are illustrated, could also be due to the presence of multi-temperature structure of the plasma, also known as the Fe-bias \citep{Simionescu2019, Mernier2022}. Next, this region is modeled with a second thermal component, with the metal abundance tied between the two \texttt{APEC} components. The temperatures and normalizations are left as free parameters; the parameters cannot be independently constrained for each region and component. Adding a second thermal component yields a better fit to the spectrum in this region. Given that the temperature of the second component is significantly higher than expected for the thermal phase of the plasma, we further investigate the nature of this X-ray emission in the following subsections. To improve statistical constraints and to leverage the similarity of plasma properties across the regions, we perform a joint spectral fit. This approach allows us to examine the nature of this component and assess whether it may be better described by a non-thermal origin, potentially providing a more accurate representation of the observed X-ray emission.

\begin{comment}
    
%
\begin{table}[htbp]
\centering
\begin{tabular}{lcc}
\hline
Parameter & (0.7-2) R$_{500c}$ & (2-4) R$_{500c}$ \\
\hline
$n_H$ & $0.037$ & $0.037$ \\
$kT_1$ (keV) & $0.90 \pm 0.03$ & $0.91 \pm 0.24$ \\
$kT_2$ (keV) & $2.07 \pm 0.50$  & $-$ \\
Abundance & $0.26 \pm 0.07$ &  $0.08 \pm 0.16$ \\
Norm$_1$ ($10^{-5}$) & $2.36 \pm 0.58$  & $0.89 \pm 1.63$  \\
Norm$_2$ ($10^{-5}$) & $1.98 \pm 0.31$  & $-$ \\
\hline
C-stat & $490.4(599)$ & $571.2 (599)$\\
\hline
\end{tabular}
\caption{Best-fit Model Parameters for the multi-temperature model. [1] NC stands for not-constrained. In  this case, metal abundance reaches the allowed upper prior\label{tab:multikT}}
\end{table}
%
\end{comment}

%
\subsection{Joint modeling of spectra beyond 0.7~R$_{500c}$}

The joint spectral analysis over the radial range beyond 0.7--4~R$_{500c}$ targets a dynamically active and volatile region in galaxy groups. This region is expected to exhibit significant thermodynamic complexity, in which the IGrM may contain a high-entropy plasma driven outward by feedback from the central black hole, alongside ongoing accretion from filamentary structures. To characterize the plasma's thermal state, we model the spectra using a range of approaches, including single- and multi-temperature thermal models, as well as a power-law component to account for potential non-thermal emission, potentially from IC processes.

\begin{figure*}
\begin{tabular}{c}
\includegraphics[width=0.48\textwidth]{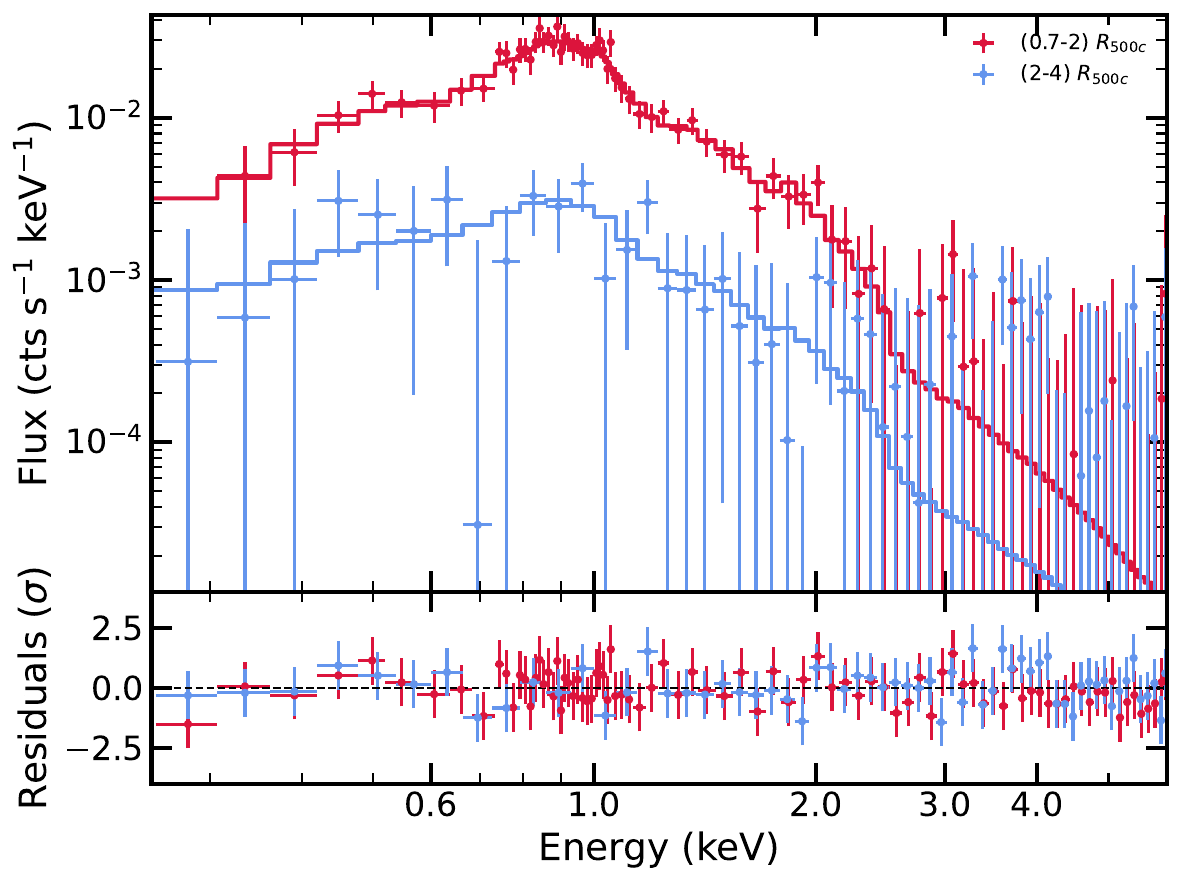} 
\includegraphics[width=0.48\textwidth]{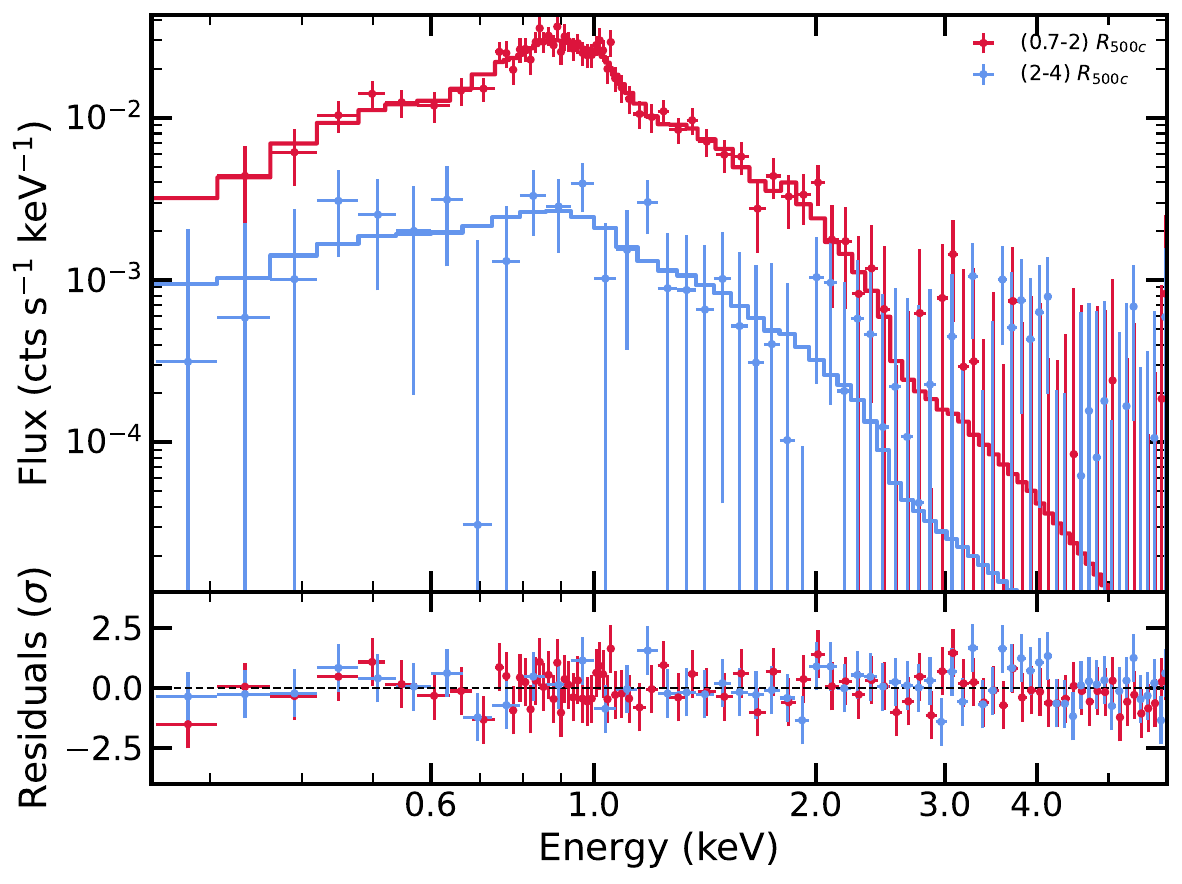} 
\end{tabular}
\caption{Joint fit of the net stacked X-ray spectra extracted from the radial ranges 0.7–2~R$_{500c}$ (red) and 2–4~R$_{500c}$ (dark blue) for the $\ntot$ galaxy groups. The best-fit model, consisting of a double-component thermal model (left panel) and a single-temperature thermal component combined with a power-law component (right panel) with free photon index and normalization, provides an improved description of the data in both the soft and hard X-ray bands, and yields robust metal abundance constraints for both regions over the single temperature models with similar goodness of fit values. \label{fig:stacked_spec_2t}}
\end{figure*}

For the joint spectral analysis, we improve statistical constraints and reduce the number of free parameters by linking the temperatures and metal abundances of the two regions, given their similarity in Regions~1 and 2, while allowing the normalizations to vary independently. The single-temperature thermal model yields consistent parameter estimates, with a temperature of \apecdsOnepriTmean~keV and a metallicity of \apecdsOnepriabundanc~A$_{\odot}$, as summarized in Table~\ref{tab:spectral_fits}. We note that the residuals in the soft and hard bands are also present in the joint fits. Interestingly, the best-fit temperature is consistent with the expected weighted mean value of $\sim 0.9$~keV of the galaxy groups in the sample, as anticipated in the infall regions around and beyond R$_{500c}$ (see Sec.~\ref{sec:tempdist} for the details). The Bayesian evidence is $\ln\mathcal{Z}=-594.2\pm 0.3$ for this model.

In reality, the thermal structure of the stacked emission from a population of sources is expected to be more complex than a single-temperature plasma, particularly in the outskirts regions where gas temperature variations and blended Fe-L emission can significantly affect the observed spectral shape. To investigate whether a broader temperature distribution provides a more physical description of the primary thermal component, we fit the spectra using the Gaussian differential emission measure (\texttt{GADEM}) model in \texttt{XSPEC}.
The best-fit mean temperature and width of the temperature distribution are \gademdsOnepriTmean~keV and \gademdsOnepriTsigma~keV, respectively. This temperature distribution is consistent with the expected temperatures of the stacked sample, as shown in Fig.~\ref{fig:sample_kT} and inferred from the single-temperature \texttt{apec} fits. The \texttt{GADEM} model yields a metallicity of \gademdsOnepriabundanc~A$_{\odot}$, higher than that obtained from the single-temperature model, suggesting that accounting for the multi-temperature structure can alleviate the Fe bias commonly present in single-temperature fits.

The best-fit normalizations of \gademdsOneprinorm$\times10^{-5}$~cm$^{-5}$ and \gademdsTwoprinorm$\times10^{-5}$~cm$^{-5}$ in Regions 1 and 2 correspond to electron densities of $(4.8\pm1.3)\times10^{-5}$~cm$^{-3}$ and $(5.5\pm2.0)\times10^{-6}$~cm$^{-3}$, respectively, assuming constant temperature and abundance within each shell. These measurements probe gas densities significantly lower than those typically accessible with previous X-ray observations \citep{Bulbul2016, Eckert2025}. The corresponding electron pressures are $P_e=1.4\pm0.5\times10^{-13}$~erg~cm$^{-3}$ and $P_e=9.2\pm4.5\times10^{-13}$~erg~cm$^{-3}$ in Regions 1 and 2, respectively. The \texttt{GADEM} model achieves a Bayesian evidence of $\ln\mathcal{Z}=-591.7\pm0.5$, corresponding to a log-Bayes factor of $\Delta\ln\mathcal{Z}=2.5\pm0.6$ and a Bayes factor of $\mathcal{K}\sim12$ relative to the single-temperature \texttt{apec} model. This provides moderate evidence that a distributed temperature structure provides a better description of the observed emission, although the current data do not require a multi-temperature model at a high significance level.

Introducing a secondary thermal component modeled with \texttt{GADEM} yields parameter values broadly consistent with those obtained from the individual regional fits (Sec.~\ref{sec:individualFits}). We first link the metal abundances of the two thermal components. In this configuration, the primary component exhibits a slightly lower temperature and a mildly enhanced abundance of \gademGademdsOnepriabundanc~A$_{\odot}$, although the value remains consistent with the single-component fit within $1\sigma$ uncertainties. The additional thermal component is characterized by a higher mean temperature of \gademGademdsOnesecTmean~keV and a temperature dispersion of \gademGademdsOnesecTsigma~keV. However, this temperature is higher than expected for gas located in the outskirts of galaxy groups at and beyond R$_{500c}$, in the infall regions. When the abundance of the secondary component is allowed to vary independently, it converges to zero and remains unconstrained. Despite these uncertainties, the inclusion of the additional component improves the fit, increasing the Bayesian evidence to $\ln\mathcal{Z}=-587.8\pm0.3$.

Motivated by the unusually high temperature and the poorly constrained abundance of the secondary component, we further explore models that include a possible non-thermal contribution. We test a model combining a thermal \texttt{GADEM} component with a power-law component to investigate a potential inverse Compton origin of the excess emission. IC scattering of CMB photons by relativistic electrons can produce X-ray emission with a power-law spectral shape and, in low-resolution spectra, can also resemble a thermal continuum with negligible line emission. We therefore consider both interpretations in our spectral modeling.

\begin{table*}[t]
\centering
\caption{Comparison of spectral fit parameters derived from BXA marginalized posteriors across different model configurations.}
\begin{tabular}{lccccc}
\hline\hline
Parameter & APEC & GADEM & GADEM + GADEM & GADEM + Brems & GADEM + PL \\
\hline
\textbf{Primary Component} & & & & & \\
$T_{\text{mean}}$ (keV) & \apecdsOnepriTmean & \gademdsOnepriTmean & \gademGademdsOnepriTmean & \gademBremsdsOnepriTmean & \gademPLdsOnepriTmean \\
$T_{\text{sigma}}$ (keV) & --- & \gademdsOnepriTsigma & \gademGademdsOnepriTsigma & \gademBremsdsOnepriTsigma & \gademPLdsOnepriTsigma \\
Metallicity ($Z_{\odot}$) & \apecdsOnepriabundanc & \gademdsOnepriabundanc & \gademGademdsOnepriabundanc & \gademBremsdsOnepriabundanc & \gademPLdsOnepriabundanc 
\\
Norm, Reg. 1 ($10^{-5}$) & \apecdsOneprinorm & \gademdsOneprinorm & \gademGademdsOneprinorm & \gademBremsdsOneprinorm & \gademPLdsOneprinorm \\
Norm, Reg. 2 ($10^{-5}$) & \apecdsTwoprinorm & \gademdsTwoprinorm & \gademGademdsTwoprinorm & \gademBremsdsTwoprinorm & \gademPLdsTwoprinorm \\
\\
\hline\\
\textbf{Secondary Component} & & & & & \\
Brem $kT$ (keV) & --- & --- & --- & \gademBremsdsOnebremsTmean & --- \\
PL Index ($\Gamma$) & --- & --- & --- & --- & \gademPLdsOneplPhoIndex \\
GADEM $T_{\text{mean, 2}}$ (keV) & --- & --- & \gademGademdsOnesecTmean & --- & --- \\
GADEM $T_{\text{sigma, 2}}$ (keV) & --- & --- & \gademGademdsOnesecTsigma & --- & --- \\
Norm, Reg. 1 ($10^{-5}$) & --- & --- & \gademGademdsOnesecnorm & \gademBremsdsOnebremsnorm & \gademPLdsOneplnorm \\
Norm, Reg. 2 ($10^{-5}$) & --- & --- & \gademGademdsTwosecnorm & \gademBremsdsTwobremsnorm & \gademPLdsTwoplnorm \\
\\
\hline
C-stat (dof) &  1157.3 (1222) & 1147.3 (1221) & 1139.2 (1217) & 1138.9 (1218) &  1139.3 (1218) \\
$\ln Z$ & $-594.2 \pm 0.3$ & $-591.7 \pm 0.5$ & $-587.8 \pm 0.3$ & $-587.2 \pm 0.3$ & $-587.8 \pm 0.3$ \\
\hline\hline
\end{tabular}
\label{tab:spectral_fits}
\end{table*}

For the model comprising a thermal \texttt{GADEM} component and a power-law component, the best-fit mean plasma temperature is \gademPLdsOnepriTmean~keV, consistent with the temperatures derived from the individual regions. The width of the temperature distribution is characterized by a median temperature and standard deviation of \gademPLdsOnepriTsigma~keV, again consistent with the expected median temperature of the sample of approximately 0.9~keV (Section~\ref{sec:tempdist}). The inferred metal abundance, \gademPLdsOnepriabundanc~A$_{\odot}$, is also physically plausible and broadly consistent with measurements of nearby galaxy groups and clusters \citep{Sun2009, Bulbul2016, Mernier2016, Urban2017, Eckert2025}.

Observations with Suzaku have enabled measurements of the thermodynamic and chemical properties of the intracluster medium out to R$_{200}$ in galaxy clusters, providing important constraints on the distribution of metals in cluster outskirts \citep{Bulbul2016, Urban2017}. Similar measurements for galaxy groups remain considerably more challenging because of their lower gas temperatures and substantially lower surface brightness, particularly beyond R$_{500}$. Among the few available studies extending to R$_{200c}$, \citet{Gastaldello2021, Sarkar2022} reported metallicities of approximately 0.2--0.3~A$_{\odot}$, adopting the \citet{Asplund2009} solar abundance scale. These values are somewhat higher than the abundances inferred in the present work.

Allowing both parameters of the power-law component to vary yields a photon index of \gademPLdsOneplPhoIndex\ and normalizations of (\gademPLdsOneplnorm)$\times10^{-5}$~cm$^{-2}$~s$^{-1}$~keV$^{-1}$ and (\gademPLdsTwoplnorm)$\times10^{-5}$~cm$^{-2}$~s$^{-1}$~keV$^{-1}$ at 1~keV for the (0.7--2)~R$_{500c}$ and (2--4)~R$_{500c}$ regions, respectively. The power-law component is detected only marginally, with a significance of approximately $2\sigma$. Nevertheless, the model provides an adequate description of the data, with a Bayesian evidence of $\ln\mathcal{Z} = -587.8 \pm 0.3$, comparable to that obtained for the other model combinations considered.

In the 0.3--6~keV band, the thermal component has fluxes of $5.2\times10^{-14}$~erg~cm$^{-2}$~s$^{-1}$ and $3.4\times10^{-15}$~erg~cm$^{-2}$~s$^{-1}$ in Regions~1 and 2, respectively. The corresponding power-law fluxes are $2.4\times10^{-14}$~erg~cm$^{-2}$~s$^{-1}$ and $6.7\times10^{-15}$~erg~cm$^{-2}$~s$^{-1}$, which, within this interpretation, are attributed to non-thermal emission. Within this model, the best-fit power-law component contributes approximately 50\% of the thermal flux in both regions, indicating that, if present, the non-thermal emission represents a substantial fraction of the total X-ray emission. The inferred non-thermal flux is approximately a factor of 3--5 lower than the inverse-Compton flux reported for the bright cores of nearby galaxy groups by \citet{Mernier2023}.

A non-thermal IC component produced by relativistic electrons can exhibit a continuum spectrum that resembles thermal emission but lacks the characteristic emission lines of a thermal plasma \citep{Hopkins2026}. To investigate this possibility, we model the spectra with a thermal \texttt{GADEM} component together with an additional thermal Bremsstrahlung component. The best-fit mean temperature of the primary \texttt{GADEM} component is \gademBremsdsOnepriTmean~keV, with a Gaussian width of \gademBremsdsOnepriTsigma~keV, consistent with the values obtained for the other models considered. The corresponding metal abundance, \gademBremsdsOnepriabundanc~A$_{\odot}$, is likewise consistent with the previous fits. The Bremsstrahlung component has a best-fit temperature of \gademBremsdsOnebremsTmean~keV, which is, as expected, slightly higher than that of the primary thermal component. This model combination provides an equally good description of the data, yielding a Bayesian evidence of $\ln\mathcal{Z} = -587.2 \pm 0.3$, statistically indistinguishable from that obtained for the thermal plus power-law model.

We also explored a three-temperature thermal model; however, the parameters of the additional component are poorly constrained. None of these additional components leads to a significant improvement in the overall fit. We therefore find no compelling evidence for the presence of an additional thermal component in the infall regions of galaxy groups.
The models, including an additional secondary non-thermal (both with power-law and Bremsstrahlung components) or thermal component, yield similar Bayesian evidences, with ($\ln\mathcal{Z} \sim -587$). The resulting best-fit model and data for these two cases are shown in Figure~\ref{fig:stacked_spec_2t}. This represents a significant improvement compared to the single-component thermal model, which has ($\ln\mathcal{Z} = -591.7$). Within the Bayesian framework, the Bayes factor ($\mathcal{K}$), defined as the ratio of the Bayesian evidence ($\mathcal{Z}$) for the two compared models, quantifies the relative support for one model over another. The resulting evidence difference ($\ln \mathcal{K} = 5$) corresponds to strong evidence in favor of the models with an additional component according to the scale of \citet{Kass1995}. These comparisons indicate that an additional emission component, either thermal or non-thermal, is significantly favored in the galaxy group's infall regions. However, the current data favor the presence of an additional spectral component but do not distinguish between thermal and non-thermal interpretations, nor identify its underlying physical origin.

\section{Interpretation and Systematics}
\label{sec:discussion}
Our analysis reveals the presence of an additional emission component beyond the thermal plasma, which can be interpreted either as a secondary thermal component or as a non-thermal contribution in the infall regions of a stacked sample of galaxy groups extending around and beyond R$_{500c}$. If interpreted as thermal emission, the additional spectral component corresponds to a plasma temperature of approximately $\sim 3.6$~keV over radii spanning 0.7~R$_{500c}$ to 4~R$_{500c}$ in galaxy groups.
Notably, when the metal abundance is allowed to vary freely, it is uncontrained. This is unexpected, as gas in the infall regions beyond the Virial radius is generally expected to be diffuse and cooler than IGrM, with temperatures typically below $\sim 1$~keV.

The origin of such high-temperature plasma in these outer regions is not straightforward to interpret. One possibility is that the emission arises from an extended IGrM bound to the group halo, tracing the outer reaches of the hot gaseous atmosphere. However, maintaining a $\sim 3.6$~keV component out to several virial radii is difficult to reconcile with a standard, relaxed IGrM scenario and would likely require strong mechanical feedback from the central black hole operating over distances of $\sim 0.3$–$2$~Mpc. Megaparsec-scale jets are indeed observed in galaxy groups \citep{Oei2024}, and similarly strong feedback is suggested by kSZ measurements of high-redshift, optically selected galaxy groups \citep{Siegel2025, Hadzhiyska2024, Popesso2024}. Given that this hotter component contributes significantly (approximately half of the thermal contribution in flux) to the total emissivity and flux than the cooler $\sim 1$~keV plasma, a substantial fraction of the gas would need to be expelled, implying the presence of powerful, large-scale AGN jets.

Another interpretation is that the observed component traces gas heated by
Virial shocks during large-scale structure formation. 
The X-ray emission in the infall and Virial regions of haloes arises from gas associated
with the surrounding large-scale structure. 
Recently, observations with eROSITA have revealed
diffuse X-ray emission associated with filamentary structures linking clusters
and groups, with temperatures close to the upper end of the expected WHIM
range \citep{Reiprich2021, Veronica2024, Zhang2024, Zhang2025}. 
In regions where filament gas interacts
with the gravitational potential of galaxy groups, gravitational compression
and shock heating may increase the temperature beyond the canonical temperature range of 0.1-1~keV. As
matter flows along filaments toward the gravitational potential wells of groups
and clusters, it is expected to pass through Virial shock fronts with several Mach numbers
located near or beyond the virial radius. These shocks can heat the infalling
gas to temperatures of several keV, producing extended and faint X-ray emission
in the outskirts of halos \citep{Molnar2009, Keshet2020}. Recent analyses of
X-ray and radio catalogs have reported statistical evidence for sources associated
with virial shocks around clusters and groups at radii of roughly
$\sim2\,R_{500c}$, consistent with hot gas accreting from the cosmic web
\citep{Ilani2024}. If the emission detected here traces shocked gas, it may provide evidence consistent with ongoing accretion. However, the measured temperature of this component ($\sim$3.6~keV) is significantly higher than the virial shock temperature expected for galaxy groups with masses of this sample.

An alternative interpretation is that part of the additional emission originates from non-thermal processes, specifically IC scattering of low-energy CMB photons by relativistic electrons \citep{Ruszkowski2023}. Merger-driven shocks in the IGrM can accelerate electrons via diffusive shock acceleration, producing a non-thermal population extending to GeV–TeV energies. Accretion-shock-accelerated cosmic rays are expected to have a DSA-like power-law spectrum, reflecting their origin via shock acceleration. Cosmic rays produced by the central AGN can also be transported to large cluster radii and accumulate over many Gyrs. Low-energy electrons have long cooling times and can propagate over large distances \citep{Quataert2025}, where they may be further accelerated to higher energies by turbulence in the cluster outskirts.
These relativistic electrons upscatter CMB photons through IC scattering, yielding X-ray emission \citep{Hopkins2026}.  These cosmic rays are the oldest and therefore produce an X-ray spectrum that more closely resembles thermal bremsstrahlung, but without the accompanying emission lines. IC emission traces the distribution of cosmic-ray electrons and provides a complementary diagnostic to radio synchrotron observations, as the same electron population produces both IC X-rays and synchrotron radio emission, depending on the ambient magnetic field strength \citep{Rephaeli2008, Brunetti2014}.
In galaxy group infall regions, IC emission is expected to dominate over synchrotron emission unless the magnetic field strength is several $\mu G$. This IC X-ray emission is faint but may dominate over thermal emission in low-density regions where the IGrM is underdense. This is particularly relevant beyond R$_{500c}$, where the thermal surface brightness declines steeply while the CR distribution is much less steep \citep{Quataert2025}. In this scenario, the observed X-ray flux directly probes the relativistic electron population in the outskirts of galaxy groups \citep{Petrosian2001}.

In addition to cosmic rays accelerated by accretion shocks and those originating from the central AGN, galaxy groups (with masses $>10^{13}$~M$_{sun}$) can have a significant, or even dominant, cosmic ray contribution in their outskirts from the combined supernovae and AGN activity of satellite and neighboring galaxies. These cosmic rays are expected to have an intermediate spectral shape, less aged than cosmic rays from the central AGN but older than freshly accelerated shock cosmic rays. Because the total stellar mass, black hole mass, and AGN power in groups are often dominated by the satellite population rather than the central galaxy, these satellites can inject substantial cosmic ray energy, making them an important contributor to diffuse emission at $\sim$Mpc scales.

If the hard component has an IC origin, the emission intensity scales with the relativistic electron density and the CMB strength. Meanwhile, the same population of relativistic electrons emits synchrotron radiation in the radio wavelengths, whose intensity scales with the relativistic electron density and the intergalactic magnetic field strength. Thus, the best-fit X-ray intensity provides insight into the relativistic electron density, which, in turn, allows the magnetic field strength to be inferred from the radio intensity at GHz wavelengths. \citet{Vernstrom2023} reported detections of stacked radio emission between luminous red galaxy pairs. Though their stacking regions are not exactly the same as those in this work, the intergalactic space between luminous red galaxies covers $>r_{500}$ outskirts regions of galaxy groups, and therefore we use the reported radio properties therein for magnetic field strength estimation. 

We used the package \texttt{Naima} \citep{Zabalza2015} to calculate synchrotron and IC emission. For this plausible IC emission, the integrated flux density at 1~keV reported in Table \ref{tab:spectral_fits} is approximately $3\times10^{-6}$~cts~s$^{-1}$~keV$^{-1}$. Meanwhile, the CMB temperature and energy density at $z=0.15$ are $3.13$~K and $7.3\times10^{-13}$~erg~cm$^{-3}$, respectively. The photon index from X-ray fits has a large uncertainty, but agrees with the radio spectral index reported by \citet{Vernstrom2023} within $1\sigma$ if assuming the two types of emission originate from the same population of relativistic electrons. Therefore, we use $\Gamma=2$, which corresponds to spectral index $\alpha=1$ reported by \citet{Vernstrom2023} for this calculation. We found that a power law relativistic electron spectrum with a normalization of $1.7\times10^{40}$~eV$^{-1}$ at 1~GeV can reproduce the observed intensity of the plausible X-ray IC emission. We use the same relativistic electron spectrum to calculate the synchrotron emission. \citet{Vernstrom2023} reported an averaged brightness temperature at 100~MHz is about 0.4~K. We used the sample median 0.7--4~$r_\mathrm{500}$ solid angle to convert the brightness temperature to the integrated radio flux. To match the reported radio intensity, the magnetic field strength is approximately 0.15~$\mu$G. One could argue that the plausible IC emission shows a large gradient across the two analysis regions, and the reported radio emission in \citet{Vernstrom2023} is more likely to be associated with the extremely faint X-ray emission in the 2--4~$R_{500c}$ region. Given that the CMB strength is fixed, a lower X-ray intensity suggests a lower relativistic electron abundance, which requires a stronger magnetic field to reproduce the same radio intensity. From our analysis, the outer-region flux is about a factor of 5 lower than the inner-region flux, suggesting the inferred magnetic field strength could be a factor of $\sqrt{5}$ stronger. Our estimation is slightly in tension with the reported intergalactic magnetic field strength $B<0.06$~$\mu$G by \citet{Vernstrom2021}. We argue that \citet{Vernstrom2021} assumed all stacked galaxy-pair X-ray signals as IC emission, which significantly overestimates the intergalactic X-ray IC strength, since our previous cosmic filament spectral stacking analysis shows that a significant amount of the intergalactic X-ray emission is likely from $T<10^7$~K thermal gas, which has a soft spectral shape \citep{Zhang2024}. Therefore, \citet{Vernstrom2021} B-field estimation using IC condition would be closer to our results if they assume the IC is only a small fraction of the stacked total X-ray signal. 
Under the assumption that the excess is dominated by inverse Compton emission, our modeling suggests magnetic field strengths of order $B\gtrsim0.1\mu$G, although this estimate is subject to systematic uncertainties. Moreover, because we cannot guarantee that all the flux of the detected hard component is from IC, the true magnetic field strength could be in the wide sub-$\mu$G range.

\subsection{Caveats and Systematics}

The detection of faint spectral features with eROSITA is fundamentally constrained by systematic uncertainties in instrument calibration, background modeling, and spectral response, despite the unprecedented statistical precision we can achieve by stacking a large number of survey observations. Systematic effects can arise from uncertainties in the effective area, energy redistribution, and detector gain, which influence the reconstructed photon energies and the inferred fluxes of weak spectral components. In low-surface-brightness regions, such as galaxy groups and their infall regions near Virial radii, minor variations in calibration data can either mimic or obscure subtle spectral features, potentially biasing derived physical parameters. We reduce the impact of calibration and background uncertainties by implementing several key analysis strategies, as described in this section.

To mitigate the calibration uncertainties, we blue-shift the spectra to the rest frame prior to stacking, which can smooth out subtle detector features \citep[e.g.,][]{Bulbul2014, Zhang2024}, thereby reducing their impact on the stacked spectrum. Because the detected emission is primarily continuum rather than line, calibration uncertainties across the entire soft band are unlikely to significantly bias the results. Additionally, the effect of the absolute calibration differences between different X-ray telescopes and eROSITA is explored using early eROSITA data \citep{Liu2023, Migkas2024, Ramos-Ceja2025}. In the softest energy band, the most relevant for emission from low-mass galaxy groups, the observed calibration differences are minimal, showing remarkable agreement within a few percent with both \textit{Chandra} and \textit{XMM-Newton} observations. If indeed present, the emission detected here should also be detected with \textit{XMM-Newton} if a similar signal-to-noise ratio and background level are reached.

Another effect that could bias the results in the soft band ($<1$~keV) is the optical light leak in some eROSITA Telescope Modules (TM) \citep[see][]{Predehl2021, Brunner2022, Merloni2024}. The absence of the optical filter in eROSITA  TM5 and TM7 introduces excess emission below 1~keV, contaminating the soft band. To avoid this contamination, we restrict stacking to Telescope Modules 1, 2, 3, 4, and 6, which are equipped with optical filters and thus unaffected by the light leak. 

Careful treatment of background modeling and spatial correlations is 
required to determine whether the detected emission traces genuine hot gas in the outskirts of galaxy groups or arises from the projected background contributions. The possibility that the signal is influenced by projection effects or unresolved background sources should also be considered. We remove all the eRASS:5-detected points and extended sources from our analysis. However, given the relatively faint emission expected at large radii, contributions from projected background components must be carefully considered. In particular, the cosmic X-ray background, which is largely produced by unresolved active galactic nuclei, may contribute to the measured spectrum \citep{Hickox2006, Moretti2012}. In some cases, unresolved structures or faint background halos projected along the line of sight may produce emission that can be approximated by a thermal component at a few keV.

Additionally, the CXB, which includes emission from unresolved active galactic nuclei, can affect spectral fits at low surface brightness levels. Consequently, accurate background modeling and consideration of spatial correlations are essential to assess whether the detected emission arises from genuine hot gas in the outskirts of galaxy groups. To assess this effect, we measure the normalized count rate in the hard X-ray band (6–10~keV) for both the source and background regions. In this energy range, galaxy groups are not expected to contribute significant emission; thus, the observed signal is dominated by the instrumental background and the cosmic X-ray background. We find that the count rate ratio between the background and source regions remains consistent across all stacked observations, indicating that the hard X-ray background component has been successfully subtracted. 

In this study, evaluating the impact of the foreground is especially important, as the detected signal appears as a continuum in the soft X-ray band. However, eROSITA has a finite detection limit for both point-like and extended sources. Diffuse sources below this threshold may remain undetected, which is particularly relevant in the infall regions of galaxy groups embedded within the large-scale structure \citep{Malavasi2020, Marini2025}. Local background regions are selected around the galaxy groups at radii of 4–6~R$_{500c}$, covering a large solid angle. These regions sample the soft band ($<2$~keV), dominated by Milky Way foreground emission. The smooth background spectra in the softest band ($<1$~keV), where Milky Way emission and solar wind charge exchange dominate, confirm that foreground contributions are properly accounted for. The contribution from O~VII and O~VIII solar wind charge exchange is also effectively removed, as no excess emission is detected at $\sim 0.56$~keV or $\sim0.65$~keV, where strong lines from this process are typically expected. Spatial variations in column density across the large survey area can introduce systematic uncertainties in the normalization, up to a factor of $\sim \sqrt{2}$, in the eROSITA survey (White et al., 2026, in prep.). Therefore, local foreground is unlikely to make up the observed signal in the outskirts of galaxy groups stacked here.

The finite point spread function (PSF) of eROSITA, with a characteristic size of $\sim 30^{\arcsec}$, limits our ability to fully remove bright point sources \citep{Brunner2022}. In particular, the PSF wings can contribute to the measured emission, especially in the hard band, if not properly accounted for. In this work, we adopt a variable-radius masking approach for point sources, in which the exclusion region is defined based on the local background-to-source surface brightness reaching 20\%. Specifically, we determine the radius at which the source signal declines to the background level \citep{Zhang2025}. A similar issue arises for extended sources; aggressively removing nearby large-scale structures can eliminate a significant fraction of the genuine source signal, while insufficient masking can lead to contamination \citep{Liu2022, Bulbul2022}. To mitigate this effect, we also implement a dynamic exclusion radius for extended sources, determined by fitting their surface brightness profiles and identifying the radius at which the emission reaches the background level (Ding et al., in prep.). This method should successfully remove the extended emission from diffuse sources. Indeed, during visual inspections, we did not detect any residual contaminating signal in the images.

\section{Conclusions}
\label{sec:conclusions}
The faint infall regions of galaxy groups and clusters have historically remained poorly explored due to observational and instrumental limitations. With the launch of eROSITA, this situation has changed substantially, enabling an unprecedented increase in statistical power. Owing to its high sensitivity in the soft X-ray band and wide field of view, eROSITA is uniquely well-suited for probing the properties of IGrM plasma at large radii through stacking analyses. Recent studies have reported detections of diffuse warm-hot gas in the infall regions of galaxy clusters, well beyond the Virial radius, and in cosmic filaments spanning 20--100~Mpc, tracing both the intracluster and intergalactic media. These results demonstrate the feasibility of detecting diffuse, low-density, low-temperature X-ray emission in the large-scale structures with eROSITA \citep{Zhang2024, Zhang2025}.

In this work, we present the first measurements of the spectroscopic properties of the IGrM plasma in group-mass halos (M$_{tot}\,<1\times10^{14}$~M$_{sun}$), based on a stacking analysis reaching beyond the virial radius ($\sim 2$~R$_{200m}$=2.2~Mpc). The sample used here is drawn from the primary catalog in the first eROSITA All-Sky Survey \cite{Bulbul2024}, and is uniformly selected with minimal contamination. Specifically, we stack spectra of \ntot nearby galaxy groups detected in the first survey, extracted from the deepest eROSITA data, from the cumulative five consecutive surveys. In this work, we closely examine the spectroscopic properties of the infall region around R$_{500c}$ (with a sample median radius of 567~kpc) and beyond out to 2.2~Mpc. Through stacking, we reach unprecedented levels that were previously unattainable. The exposure time in the source region amounts to 3.5~Ms, yielding a total of 42,000 counts after background subtraction in the region (0.7-2)~R$_{500c}$. For region (2-4)~R$_{500c}$, the corresponding exposure time is 5.1~Ms, with 15,400 net source counts. The total background exposure reaches 7.1~Ms.

When modeling the stacked spectra, a thermal model (\texttt{GADEM}) provides a satisfactory overall fit. In the radial range (0.7--4)~R$_{500c}$, we obtain a plasma temperature distribution with a weighted mean of \gademGademdsOnepriTmean~keV with a width of \gademGademdsOnepriTsigma~keV. The temperature is consistent with the expected weighted mean value of $\sim 0.9$~keV for the galaxy groups in the sample, as anticipated in the infall regions around and beyond R$_{500c}$. The metal abundance of \gademGademdsOnepriabundanc~A$_{sun}$ is consistent with the expectation from numerical simulations at this radius. Assuming a constant temperature and abundance across the annuli, from spectroscopy we measure the electron number density to be $(4.8\pm 1.3) \times 10^{-5}$~cm$^{-3}$ and $(5.5 \pm 2.0) \times10^{-6}$~cm$^{-3}$ in regions (0.7--2)~R$_{500c}$ and (2--4)~R$_{500c}$, respectively. The corresponding electron pressures are $P_e=1.4\pm0.5\times10^{-13}$~erg~cm$^{-3}$ and $P_e=9.2\pm4.5\times10^{-13}$~erg~cm$^{-3}$ in Regions 1 and 2, respectively.

The present data favor the presence of an additional spectral component beyond a single thermal plasma. Both a hotter thermal component and a non-thermal inverse Compton interpretation provide comparably good descriptions of the data, and higher spectral resolution or complementary multi-wavelength observations will be required to distinguish between these scenarios.
Introducing a second thermal component in the joint fit to both regions yields a significantly higher mean temperature of \gademGademdsOnesecTmean~keV, with a Gaussian width of \gademGademdsOnesecTsigma~keV and an unconstrained metal abundance. Such a solution is difficult to reconcile with the expected thermodynamic properties of galaxy-group infall regions, indicating that the additional component is unlikely to have a purely thermal origin.

An alternative interpretation is that the excess emission arises from IC scattering of relativistic electrons. This emission would be expected to appear as either a power-law component or an emission-line-free continuum that can resemble thermal emission, while contributing preferentially at higher X-ray energies. Although the detection remains tentative, the Bayesian model comparison ($\ln \mathcal{K}=\sim5$) provides strong evidence in favor of including an additional spectral component.

Modeling the excess with a power law yields a photon index of \gademPLdsTwoplPhoIndex, consistent with values inferred for relativistic electron populations in galaxy groups from radio observations \citep{Bagchi2009}. For the power-law model, we measure fluxes of $2.5 \times 10^{-15}$ and $6.7 \times 10^{-16}$~erg~cm$^{-2}$~s$^{-1}$ in the two regions, respectively. These values are approximately a factor of three to five lower than the IC flux recently reported in the core of a galaxy group by \citet{Mernier2023}. The inferred non-thermal component contributes approximately 30\% of the thermal X-ray flux in both regions, implying that, if confirmed, inverse Compton emission could represent a substantial fraction of the total X-ray emission from the infall regions. Assuming the hard X-ray excess arises from IC scattering by the same population of relativistic electrons responsible for the diffuse radio emission, we estimate a magnetic field strength of $B \gtrsim 0.1,\mu$G in this region. This value is broadly consistent with expectations for low-density large-scale structures and suggests that previous lower estimates may have underestimated the thermal contribution to the stacked X-ray emission. Given the current uncertainties and the possibility that only part of the hard X-ray component is of IC origin, the true magnetic field strength is likely to lie within the sub-$\mu$G regime. If the excess emission is of non-thermal origin, arising from inverse Compton scattering, it would likely be associated with the low-density infall regions of the system, where accretion shocks and particle acceleration are expected to be most efficient. Detecting such emission would provide direct constraints on the population of relativistic electrons and the strength of the ambient magnetic field, yielding valuable insight into the interplay between cosmic rays, large-scale structure formation, and the thermodynamic evolution of the intragroup medium. Although eROSITA provides excellent statistical sensitivity, its limited effective area above $\sim3$~keV restricts our ability to distinguish between a secondary thermal component and a non-thermal origin, such as inverse Compton emission.

Interpreting the detected signal is not trivial in these violent regions of galaxy groups, as we cannot distinguish between thermal and non-thermal origins using spectroscopy alone. If thermal in origin, the emission implies a plasma that contributes significantly to the overall X-ray emissivity, exceeding that expected from the $\sim1$~keV component. A high-temperature plasma at $\sim3.6$~keV could, in principle, be produced by AGN feedback, where powerful Mpc-scale jets heat the surrounding medium by dispersing the plasma from inner regions. The cosmic ray density in the outskirts can be dominated by the collective cosmic rays accelerated from SNe and AGN in all the satellite and neighboring galaxies in the group. However, such extended jets are rarely observed in galaxy groups, with only a small number of confirmed cases, making it difficult to generalize this scenario to a stacked sample \citep{Oei2024}.
An alternative thermal explanation is shock-heated gas associated with ongoing accretion. In this case, the emission may trace gas heated by large-scale structure formation processes, including accretion shocks, turbulence from minor mergers, and interactions with filamentary structures in the cosmic web. As gas falls into the potential well from the surrounding large-scale environment, it may be heated to temperatures of a few keV, producing features consistent with accretion-driven shock heating. However, the measured $\sim3.6$~keV temperature remains high for the expected temperature of the shock-heated plasma in the infall regions of low-mass galaxy groups.

Although observational challenges remain due to the relative weakness of this signal compared to thermal X-ray emission, we control instrumental calibration and background-related systematics by improving the robustness of these measurements. In particular, the smoothing achieved through blueshifting and stacking over large spatial regions helps mitigate some of these uncertainties, and the adopted variable exclusion radius for both point and extended sources. The stacked X-ray spectra of galaxy groups detected in deeper eROSITA surveys, such as eRASS:5, will achieve approximately a factor of two higher detection significance, enabling a more robust assessment of the nature of the detected emission. Future investigations of infall regions in faint galaxy groups require enhanced hard X-ray sensitivity and larger collecting areas. These capabilities are expected from next-generation missions such as newAthena, as well as from deep, targeted observations \citep{Nandra2013, Cruise2025}, which will be crucial for disentangling thermal and non-thermal components in these regimes.

\begin{acknowledgement}
The authors thank Dan Wik and Congyao Zhang for insightful discussions on the interpretation of the results. E. Bulbul, E. Artis, S. Zelmer, and X. Zhang acknowledge financial support from the European Research Council (ERC) Consolidator Grant under the European Union’s Horizon 2020 research and innovation program (grant agreement CoG DarkQuest No 101002585). 
\\

This work is based on data from \rosi, the soft X-ray instrument aboard SRG, a joint Russian-German science mission supported by the Russian Space Agency (Roskosmos), in the interests of the Russian Academy of Sciences represented by its Space Research Institute (IKI), and the Deutsches Zentrum f{\"{u}}r Luft und Raumfahrt (DLR). The SRG spacecraft was built by Lavochkin Association (NPOL) and its subcontractors and is operated by NPOL with support from the Max Planck Institute for Extraterrestrial Physics (MPE).

\\
The development and construction of the \rosi\ X-ray instrument were led by MPE, with contributions from the Dr. Karl Remeis Observatory Bamberg \& ECAP (FAU Erlangen-Nuernberg), the University of Hamburg Observatory, the Leibniz Institute for Astrophysics Potsdam (AIP), and the Institute for Astronomy and Astrophysics of the University of T{\"{u}}bingen, with the support of DLR and the Max Planck Society. The Argelander Institute for Astronomy of the University of Bonn and the Ludwig Maximilians Universit{\"{a}}t Munich also participated in the science preparation for \rosi.

\\

The eROSITA data shown here were processed using the eSASS/NRTA software system developed by the German eROSITA consortium.

\end{acknowledgement}

\bibliographystyle{aa}
\bibliography{references}

@ARTICLE{Asplund2009,
       author = {{Asplund}, Martin and {Grevesse}, Nicolas and {Sauval}, A. Jacques and {Scott}, Pat},
        title = "{The Chemical Composition of the Sun}",
      journal = {\araa},
         year = 2009,
        month = sep,
       volume = {47},
       number = {1},
        pages = {481-522},
          doi = {10.1146/annurev.astro.46.060407.145222},
archivePrefix = {arXiv},
       eprint = {0909.0948},
 primaryClass = {astro-ph.SR},
       adsurl = {https://ui.adsabs.harvard.edu/abs/2009ARA&A..47..481A}
}

@ARTICLE{Bagchi2009,
       author = {{Bagchi}, Joydeep and {Jacob}, Joe and {Gopal-Krishna} and {Werner}, Norbert and {Wadnerkar}, Nitin and {Belapure}, Jaydeep and {Kumbharkhane}, A.~C.},
        title = "{A diffuse bubble-like radio-halo source MRC0116+111: imprint of AGN feedback in a low-mass cluster of galaxies}",
      journal = {\mnras},
         year = 2009,
        month = oct,
       volume = {399},
       number = {2},
        pages = {601-614},
          doi = {10.1111/j.1365-2966.2009.15310.x},
archivePrefix = {arXiv},
       eprint = {0907.1534},
 primaryClass = {astro-ph.CO},
       adsurl = {https://ui.adsabs.harvard.edu/abs/2009MNRAS.399..601B}
}

@ARTICLE{Bahar2022,
       author = {{Bahar}, Y. Emre and {Bulbul}, Esra and {Clerc}, Nicolas and {Ghirardini}, Vittorio and {Liu}, Ang and {Nandra}, Kirpal and {Pacaud}, Florian and {Chiu}, I. -Non and {Comparat}, Johan and {Ider-Chitham}, Jacob and {Klein}, Mathias and {Liu}, Teng and {Merloni}, Andrea and {Migkas}, Konstantinos and {Okabe}, Nobuhiro and {Ramos-Ceja}, Miriam E. and {Reiprich}, Thomas H. and {Sanders}, Jeremy S. and {Schrabback}, Tim},
        title = "{The eROSITA Final Equatorial-Depth Survey (eFEDS). X-ray properties and scaling relations of galaxy clusters and groups}",
      journal = {\aap},
         year = 2022,
        month = may,
       volume = {661},
          eid = {A7},
        pages = {A7},
          doi = {10.1051/0004-6361/202142462},
archivePrefix = {arXiv},
       eprint = {2110.09534},
 primaryClass = {astro-ph.CO},
       adsurl = {https://ui.adsabs.harvard.edu/abs/2022A&A...661A...7B}
}

@ARTICLE{Bahar2024,
       author = {{Bahar}, Y.~E. and {Bulbul}, E. and {Ghirardini}, V. and {Sanders}, J.~S. and {Zhang}, X. and {Liu}, A. and {Clerc}, N. and {Artis}, E. and {Balzer}, F. and {Biffi}, V. and {Bose}, S. and {Comparat}, J. and {Dolag}, K. and {Garrel}, C. and {Hadzhiyska}, B. and {Hern{\'a}ndez-Aguayo}, C. and {Hernquist}, L. and {Kluge}, M. and {Krippendorf}, S. and {Merloni}, A. and {Nandra}, K. and {Pakmor}, R. and {Popesso}, P. and {Ramos-Ceja}, M. and {Seppi}, R. and {Springel}, V. and {Weller}, J. and {Zelmer}, S.},
        title = "{The SRG/eROSITA All-Sky Survey: Constraints on AGN feedback in galaxy groups}",
      journal = {\aap},
         year = 2024,
        month = nov,
       volume = {691},
          eid = {A188},
        pages = {A188},
          doi = {10.1051/0004-6361/202449399},
archivePrefix = {arXiv},
       eprint = {2401.17276},
 primaryClass = {astro-ph.CO},
       adsurl = {https://ui.adsabs.harvard.edu/abs/2024A&A...691A.188B}
}

@ARTICLE{Bigwood2025,
       author = {{Bigwood}, Leah and {Bourne}, Martin A. and {Irsic}, Vid and {Amon}, Alexandra and {Sijacki}, Debora},
        title = "{The case for large-scale AGN feedback in galaxy formation simulations: insights from XFABLE}",
      journal = {arXiv e-prints},
         year = 2025,
        month = jan,
          eid = {arXiv:2501.16983},
        pages = {arXiv:2501.16983},
          doi = {10.48550/arXiv.2501.16983},
archivePrefix = {arXiv},
       eprint = {2501.16983},
 primaryClass = {astro-ph.CO},
       adsurl = {https://ui.adsabs.harvard.edu/abs/2025arXiv250116983B}
}

@ARTICLE{Buchner2014,
       author = {{Buchner}, J. and {Georgakakis}, A. and {Nandra}, K. and {Hsu}, L. and {Rangel}, C. and {Brightman}, M. and {Merloni}, A. and {Salvato}, M. and {Donley}, J. and {Kocevski}, D.},
        title = "{X-ray spectral modelling of the AGN obscuring region in the CDFS: Bayesian model selection and catalogue}",
      journal = {\aap},
         year = 2014,
        month = apr,
       volume = {564},
          eid = {A125},
        pages = {A125},
          doi = {10.1051/0004-6361/201322971},
archivePrefix = {arXiv},
       eprint = {1402.0004},
 primaryClass = {astro-ph.HE},
       adsurl = {https://ui.adsabs.harvard.edu/abs/2014A&A...564A.125B}
}

@ARTICLE{Brunetti2014,
       author = {{Brunetti}, Gianfranco and {Jones}, Thomas W.},
        title = "{Cosmic Rays in Galaxy Clusters and Their Nonthermal Emission}",
      journal = {International Journal of Modern Physics D},
         year = 2014,
        month = mar,
       volume = {23},
       number = {4},
          eid = {1430007-98},
        pages = {1430007-98},
          doi = {10.1142/S0218271814300079},
archivePrefix = {arXiv},
       eprint = {1401.7519},
 primaryClass = {astro-ph.CO},
       adsurl = {https://ui.adsabs.harvard.edu/abs/2014IJMPD..2330007B}
}

@article{Brunner2022,
	adsurl = {https://ui.adsabs.harvard.edu/abs/2022A&A...661A...1B},
	archiveprefix = {arXiv},
	author = {{Brunner}, H. and {Liu}, T. and {Lamer}, G. and {Georgakakis}, A. and {Merloni}, A. and {Brusa}, M. and {Bulbul}, E. and {Dennerl}, K. and {Friedrich}, S. and {Liu}, A. and {Maitra}, C. and {Nandra}, K. and {Ramos-Ceja}, M.~E. and {Sanders}, J.~S. and {Stewart}, I.~M. and {Boller}, T. and {Buchner}, J. and {Clerc}, N. and {Comparat}, J. and {Dwelly}, T. and {Eckert}, D. and {Finoguenov}, A. and {Freyberg}, M. and {Ghirardini}, V. and {Gueguen}, A. and {Haberl}, F. and {Kreykenbohm}, I. and {Krumpe}, M. and {Osterhage}, S. and {Pacaud}, F. and {Predehl}, P. and {Reiprich}, T.~H. and {Robrade}, J. and {Salvato}, M. and {Santangelo}, A. and {Schrabback}, T. and {Schwope}, A. and {Wilms}, J.},
	doi = {10.1051/0004-6361/202141266},
	eid = {A1},
	eprint = {2106.14517},
	journal = {\aap},
	month = may,
	pages = {A1},
	primaryclass = {astro-ph.HE},
	title = {{The eROSITA Final Equatorial Depth Survey (eFEDS). X-ray catalogue}},
	volume = {661},
	year = 2022}

@ARTICLE{Bulbul2014,
       author = {{Bulbul}, Esra and {Markevitch}, Maxim and {Foster}, Adam and {Smith}, Randall K. and {Loewenstein}, Michael and {Randall}, Scott W.},
        title = "{Detection of an Unidentified Emission Line in the Stacked X-Ray Spectrum of Galaxy Clusters}",
      journal = {\apj},
         year = 2014,
        month = jul,
       volume = {789},
       number = {1},
          eid = {13},
        pages = {13},
          doi = {10.1088/0004-637X/789/1/13},
archivePrefix = {arXiv},
       eprint = {1402.2301},
 primaryClass = {astro-ph.CO},
       adsurl = {https://ui.adsabs.harvard.edu/abs/2014ApJ...789...13B}
}

@ARTICLE{Bulbul2016,
       author = {{Bulbul}, Esra and {Randall}, Scott W. and {Bayliss}, Matthew and {Miller}, Eric and {Andrade-Santos}, Felipe and {Johnson}, Ryan and {Bautz}, Mark and {Blanton}, Elizabeth L. and {Forman}, William R. and {Jones}, Christine and {Paterno-Mahler}, Rachel and {Murray}, Stephen S. and {Sarazin}, Craig L. and {Smith}, Randall K. and {Ezer}, Cemile},
        title = "{Probing the Outskirts of the Early-Stage Galaxy Cluster Merger A1750}",
      journal = {\apj},
         year = 2016,
        month = feb,
       volume = {818},
       number = {2},
          eid = {131},
        pages = {131},
          doi = {10.3847/0004-637X/818/2/131},
archivePrefix = {arXiv},
       eprint = {1510.00017},
 primaryClass = {astro-ph.CO},
       adsurl = {https://ui.adsabs.harvard.edu/abs/2016ApJ...818..131B}
}

@ARTICLE{Bulbul2022,
       author = {{Bulbul}, E. and {Liu}, A. and {Pasini}, T. and {Comparat}, J. and {Hoang}, D.~N. and {Klein}, M. and {Ghirardini}, V. and {Salvato}, M. and {Merloni}, A. and {Seppi}, R. and {Wolf}, J. and {Anderson}, S.~F. and {Bahar}, Y.~E. and {Brusa}, M. and {Br{\"u}ggen}, M. and {Buchner}, J. and {Dwelly}, T. and {Ibarra-Medel}, H. and {Ider Chitham}, J. and {Liu}, T. and {Nandra}, K. and {Ramos-Ceja}, M.~E. and {Sanders}, J.~S. and {Shen}, Y.},
        title = "{The eROSITA Final Equatorial-Depth Survey (eFEDS). Galaxy clusters and groups in disguise}",
      journal = {\aap},
         year = 2022,
        month = may,
       volume = {661},
          eid = {A10},
        pages = {A10},
          doi = {10.1051/0004-6361/202142460},
archivePrefix = {arXiv},
       eprint = {2110.09544},
 primaryClass = {astro-ph.GA},
       adsurl = {https://ui.adsabs.harvard.edu/abs/2022A&A...661A..10B}
}

@ARTICLE{Bulbul2024,
       author = {{Bulbul}, E. and {Liu}, A. and {Kluge}, M. and {Zhang}, X. and {Sanders}, J.~S. and {Bahar}, Y.~E. and {Ghirardini}, V. and {Artis}, E. and {Seppi}, R. and {Garrel}, C. and {Ramos-Ceja}, M.~E. and {Comparat}, J. and {Balzer}, F. and {B{\"o}ckmann}, K. and {Br{\"u}ggen}, M. and {Clerc}, N. and {Dennerl}, K. and {Dolag}, K. and {Freyberg}, M. and {Grandis}, S. and {Gruen}, D. and {Kleinebreil}, F. and {Krippendorf}, S. and {Lamer}, G. and {Merloni}, A. and {Migkas}, K. and {Nandra}, K. and {Pacaud}, F. and {Predehl}, P. and {Reiprich}, T.~H. and {Schrabback}, T. and {Veronica}, A. and {Weller}, J. and {Zelmer}, S.},
        title = "{The SRG/eROSITA All-Sky Survey. The first catalog of galaxy clusters and groups in the Western Galactic Hemisphere}",
      journal = {\aap},
         year = 2024,
        month = may,
       volume = {685},
          eid = {A106},
        pages = {A106},
          doi = {10.1051/0004-6361/202348264},
archivePrefix = {arXiv},
       eprint = {2402.08452},
 primaryClass = {astro-ph.CO},
       adsurl = {https://ui.adsabs.harvard.edu/abs/2024A&A...685A.106B}
}

@ARTICLE{Chiu2025,
       author = {{Chiu}, I-Non and {Ghirardini}, Vittorio and {Grandis}, Sebastian and {Okabe}, Nobuhiro and {Artis}, Emmanuel and {Bulbul}, Esra and {Bahar}, Y. Emre and {Balzer}, Fabian and {Clerc}, Nicolas and {Comparat}, Johan and {Hsieh}, Bau-Ching and {Kleinebreil}, Florian and {Kluge}, Matthias and {Liu}, Ang and {Monteiro-Oliveira}, Rogerio and {Oguri}, Masamune and {Pacaud}, Florian and {Ramos Ceja}, Miriam and {Reiprich}, H. Thomas and {Sanders}, Jeremy and {Schrabback}, Tim and {Seppi}, Riccardo and {Sommer}, Martin and {Tam}, Sut-Ieng and {Umetsu}, Keiichi and {Zhang}, Xiaoyuan},
        title = "{The SRG/eROSITA All-Sky Survey. The Weak-Lensing Mass Calibration and the Stellar Mass-to-Halo Mass Relation from the Hyper Suprime-Cam Subaru Strategic Program}",
      journal = {arXiv e-prints},
         year = 2025,
        month = apr,
          eid = {arXiv:2504.01076},
        pages = {arXiv:2504.01076},
          doi = {10.48550/arXiv.2504.01076},
archivePrefix = {arXiv},
       eprint = {2504.01076},
 primaryClass = {astro-ph.CO},
       adsurl = {https://ui.adsabs.harvard.edu/abs/2025arXiv250401076C}
}

@ARTICLE{Cruise2025,
       author = {{Cruise}, Mike and {Guainazzi}, Matteo and {Aird}, James and {Carrera}, Francisco J. and {Costantini}, Elisa and {Corrales}, Lia and {Dauser}, Thomas and {Eckert}, Dominique and {Gastaldello}, Fabio and {Matsumoto}, Hironori and {Osten}, Rachel and {Petrucci}, Pierre-Olivier and {Porquet}, Delphine and {Pratt}, Gabriel W. and {Rea}, Nanda and {Reiprich}, Thomas H. and {Simionescu}, Aurora and {Spiga}, Daniele and {Troja}, Eleonora},
        title = "{The NewAthena mission concept in the context of the next decade of X-ray astronomy}",
      journal = {Nature Astronomy},
         year = 2025,
        month = jan,
       volume = {9},
        pages = {36-44},
          doi = {10.1038/s41550-024-02416-3},
archivePrefix = {arXiv},
       eprint = {2501.03100},
 primaryClass = {astro-ph.IM},
       adsurl = {https://ui.adsabs.harvard.edu/abs/2025NatAs...9...36C}
}

@INPROCEEDINGS{Dennerl2021,
       author = {{Dennerl}, Konrad and {Andritschke}, Robert and {Br{\"a}uninger}, Heinrich and {Burkert}, Wolfgang and {Burwitz}, Vadim and {Emberger}, Valentin and {Freyberg}, Michael and {Friedrich}, Peter and {Gaida}, Roland and {Granato}, Stefanie and {Hartner}, Gisela and {von Kienlin}, Andreas and {Meidinger}, Norbert and {Menz}, Benedikt and {Predehl}, Peter},
        title = "{The calibration of eROSITA on SRG}",
    booktitle = {Society of Photo-Optical Instrumentation Engineers (SPIE) Conference Series},
         year = 2021,
       editor = {{den Herder}, Jan-Willem A. and {Nikzad}, Shouleh and {Nakazawa}, Kazuhiro},
       series = {Society of Photo-Optical Instrumentation Engineers (SPIE) Conference Series},
       volume = {11444},
        month = jan,
          eid = {114444Q},
        pages = {114444Q},
          doi = {10.1117/12.2562330},
       adsurl = {https://ui.adsabs.harvard.edu/abs/2021SPIE11444E..4QD}
}

@ARTICLE{Eckert2025,
       author = {{Eckert}, D. and {Gastaldello}, F. and {Lovisari}, L. and {McGee}, S. and {Pasini}, T. and {Brienza}, M. and {Kolokythas}, K. and {O'Sullivan}, E. and {Simionescu}, A. and {Sun}, M. and {Ayromlou}, M. and {Bourne}, M.~A. and {Chen}, Y. and {Cui}, W. and {Ettori}, S. and {Finoguenov}, A. and {Gozaliasl}, G. and {Kale}, R. and {Mernier}, F. and {Oppenheimer}, B.~D. and {Schellenberger}, G. and {Seppi}, R. and {Tempel}, E.},
        title = "{Extreme AGN feedback in the fossil galaxy group SDSSTG 4436}",
      journal = {arXiv e-prints},
         year = 2025,
        month = jun,
          eid = {arXiv:2506.13907},
        pages = {arXiv:2506.13907},
          doi = {10.48550/arXiv.2506.13907},
archivePrefix = {arXiv},
       eprint = {2506.13907},
 primaryClass = {astro-ph.GA},
       adsurl = {https://ui.adsabs.harvard.edu/abs/2025arXiv250613907E}
}

@ARTICLE{Fabian2012,
       author = {{Fabian}, A.~C.},
        title = "{Observational Evidence of Active Galactic Nuclei Feedback}",
      journal = {\araa},
         year = 2012,
        month = sep,
       volume = {50},
        pages = {455-489},
          doi = {10.1146/annurev-astro-081811-125521},
archivePrefix = {arXiv},
       eprint = {1204.4114},
 primaryClass = {astro-ph.CO},
       adsurl = {https://ui.adsabs.harvard.edu/abs/2012ARA&A..50..455F}
}

@ARTICLE{Foster2012,
       author = {{Foster}, A.~R. and {Ji}, L. and {Smith}, R.~K. and {Brickhouse}, N.~S.},
        title = "{Updated Atomic Data and Calculations for X-Ray Spectroscopy}",
      journal = {\apj},
         year = 2012,
        month = sep,
       volume = {756},
       number = {2},
          eid = {128},
        pages = {128},
          doi = {10.1088/0004-637X/756/2/128},
archivePrefix = {arXiv},
       eprint = {1207.0576},
 primaryClass = {astro-ph.HE},
       adsurl = {https://ui.adsabs.harvard.edu/abs/2012ApJ...756..128F}
}

@INPROCEEDINGS{Freyberg2021,
       author = {{Freyberg}, Michael and {Perinati}, Emanuele and {Pacaud}, Florian and {Eraerds}, Tanja and {Churazov}, Eugene and {Dennerl}, Konrad and {Predehl}, Peter and {Merloni}, Andrea and {Meidinger}, Norbert and {Bulbul}, Esra and {Friedrich}, Susanne and {Gilfanov}, Marat and {Tenzer}, Chris and {Pommranz}, Christian and {Eckert}, Dominique and {Schmitt}, J{\"u}rgen and {Brusa}, Marcella and {Santangelo}, Andrea},
        title = "{SRG/eROSITA in-flight background at L2}",
    booktitle = {Society of Photo-Optical Instrumentation Engineers (SPIE) Conference Series},
         year = 2021,
       editor = {{den Herder}, Jan-Willem A. and {Nikzad}, Shouleh and {Nakazawa}, Kazuhiro},
       series = {Society of Photo-Optical Instrumentation Engineers (SPIE) Conference Series},
       volume = {11444},
        month = jan,
          eid = {114441O},
        pages = {114441O},
          doi = {10.1117/12.2562709},
       adsurl = {https://ui.adsabs.harvard.edu/abs/2021SPIE11444E..1OF}
}

@ARTICLE{Gastaldello2021,
       author = {{Gastaldello}, Fabio and {Simionescu}, Aurora and {Mernier}, Francois and {Biffi}, Veronica and {Gaspari}, Massimo and {Sato}, Kosuke and {Matsushita}, Kyoko},
        title = "{The Metal Content of the Hot Atmospheres of Galaxy Groups}",
      journal = {Universe},
         year = 2021,
        month = jun,
       volume = {7},
       number = {7},
          eid = {208},
        pages = {208},
          doi = {10.3390/universe7070208},
archivePrefix = {arXiv},
       eprint = {2106.13258},
 primaryClass = {astro-ph.CO},
       adsurl = {https://ui.adsabs.harvard.edu/abs/2021Univ....7..208G}
}

@ARTICLE{Ghirardini2024,
       author = {{Ghirardini}, V. and {Bulbul}, E. and {Artis}, E. and {Clerc}, N. and {Garrel}, C. and {Grandis}, S. and {Kluge}, M. and {Liu}, A. and {Bahar}, Y.~E. and {Balzer}, F. and {Chiu}, I. and {Comparat}, J. and {Gruen}, D. and {Kleinebreil}, F. and {Krippendorf}, S. and {Merloni}, A. and {Nandra}, K. and {Okabe}, N. and {Pacaud}, F. and {Predehl}, P. and {Ramos-Ceja}, M.~E. and {Reiprich}, T.~H. and {Sanders}, J.~S. and {Schrabback}, T. and {Seppi}, R. and {Zelmer}, S. and {Zhang}, X. and {Bornemann}, W. and {Brunner}, H. and {Burwitz}, V. and {Coutinho}, D. and {Dennerl}, K. and {Freyberg}, M. and {Friedrich}, S. and {Gaida}, R. and {Gueguen}, A. and {Haberl}, F. and {Kink}, W. and {Lamer}, G. and {Li}, X. and {Liu}, T. and {Maitra}, C. and {Meidinger}, N. and {Mueller}, S. and {Miyatake}, H. and {Miyazaki}, S. and {Robrade}, J. and {Schwope}, A. and {Stewart}, I.},
        title = "{The SRG/eROSITA all-sky survey: Cosmology constraints from cluster abundances in the western Galactic hemisphere}",
      journal = {\aap},
         year = 2024,
        month = sep,
       volume = {689},
          eid = {A298},
        pages = {A298},
          doi = {10.1051/0004-6361/202348852},
archivePrefix = {arXiv},
       eprint = {2402.08458},
 primaryClass = {astro-ph.CO},
       adsurl = {https://ui.adsabs.harvard.edu/abs/2024A&A...689A.298G}
}

@ARTICLE{Grandis2024,
       author = {{Grandis}, S. and {Ghirardini}, V. and {Bocquet}, S. and {Garrel}, C. and {Mohr}, J.~J. and {Liu}, A. and {Kluge}, M. and {Kimmig}, L. and {Reiprich}, T.~H. and {Alarcon}, A. and {Amon}, A. and {Artis}, E. and {Bahar}, Y.~E. and {Balzer}, F. and {Bechtol}, K. and {Becker}, M.~R. and {Bernstein}, G. and {Bulbul}, E. and {Campos}, A. and {Carnero Rosell}, A. and {Carrasco Kind}, M. and {Cawthon}, R. and {Chang}, C. and {Chen}, R. and {Chiu}, I. and {Choi}, A. and {Clerc}, N. and {Comparat}, J. and {Cordero}, J. and {Davis}, C. and {Derose}, J. and {Diehl}, H.~T. and {Dodelson}, S. and {Doux}, C. and {Drlica-Wagner}, A. and {Eckert}, K. and {Elvin-Poole}, J. and {Everett}, S. and {Ferte}, A. and {Gatti}, M. and {Giannini}, G. and {Giles}, P. and {Gruen}, D. and {Gruendl}, R.~A. and {Harrison}, I. and {Hartley}, W.~G. and {Herner}, K. and {Huff}, E.~M. and {Kleinebreil}, F. and {Kuropatkin}, N. and {Leget}, P.~F. and {Maccrann}, N. and {Mccullough}, J. and {Merloni}, A. and {Myles}, J. and {Nandra}, K. and {Navarro-Alsina}, A. and {Okabe}, N. and {Pacaud}, F. and {Pandey}, S. and {Prat}, J. and {Predehl}, P. and {Ramos}, M. and {Raveri}, M. and {Rollins}, R.~P. and {Roodman}, A. and {Ross}, A.~J. and {Rykoff}, E.~S. and {Sanchez}, C. and {Sanders}, J. and {Schrabback}, T. and {Secco}, L.~F. and {Seppi}, R. and {Sevilla-Noarbe}, I. and {Sheldon}, E. and {Shin}, T. and {Troxel}, M. and {Tutusaus}, I. and {Varga}, T.~N. and {Wu}, H. and {Yanny}, B. and {Yin}, B. and {Zhang}, X. and {Zhang}, Y. and {Alves}, O. and {Bhargava}, S. and {Brooks}, D. and {Burke}, D.~L. and {Carretero}, J. and {Costanzi}, M. and {da Costa}, L.~N. and {Pereira}, M.~E.~S. and {De Vicente}, J. and {Desai}, S. and {Doel}, P. and {Ferrero}, I. and {Flaugher}, B. and {Friedel}, D. and {Frieman}, J. and {Garc{\'\i}a-Bellido}, J. and {Gutierrez}, G. and {Hinton}, S.~R. and {Hollowood}, D.~L. and {Honscheid}, K. and {James}, D.~J. and {Jeffrey}, N. and {Lahav}, O. and {Lee}, S. and {Marshall}, J.~L. and {Menanteau}, F. and {Ogando}, R.~L.~C. and {Pieres}, A. and {Plazas Malag{\'o}n}, A.~A. and {Romer}, A.~K. and {Sanchez}, E. and {Schubnell}, M. and {Smith}, M. and {Suchyta}, E. and {Swanson}, M.~E.~C. and {Tarle}, G. and {Weaverdyck}, N. and {Weller}, J.},
        title = "{The SRG/eROSITA All-Sky Survey: Dark Energy Survey year 3 weak gravitational lensing by eRASS1 selected galaxy clusters}",
      journal = {\aap},
         year = 2024,
        month = jul,
       volume = {687},
          eid = {A178},
        pages = {A178},
          doi = {10.1051/0004-6361/202348615},
archivePrefix = {arXiv},
       eprint = {2402.08455},
 primaryClass = {astro-ph.CO},
       adsurl = {https://ui.adsabs.harvard.edu/abs/2024A&A...687A.178G}
}

@ARTICLE{Hadzhiyska2024,
       author = {{Hadzhiyska}, B. and {Ferraro}, S. and {Ried Guachalla}, B. and {Schaan}, E. and {Aguilar}, J. and {Battaglia}, N. and {Bond}, J.~R. and {Brooks}, D. and {Calabrese}, E. and {Choi}, S.~K. and {Claybaugh}, T. and {Coulton}, W.~R. and {Dawson}, K. and {Devlin}, M. and {Dey}, B. and {Doel}, P. and {Duivenvoorden}, A.~J. and {Dunkley}, J. and {Farren}, G.~S. and {Font-Ribera}, A. and {Forero-Romero}, J.~E. and {Gallardo}, P.~A. and {Gazta{\~n}aga}, E. and {Gontcho Gontcho}, S. and {Gralla}, M. and {Le Guillou}, L. and {Gutierrez}, G. and {Guy}, J. and {Hill}, J.~C. and {Hlo{\v{z}}ek}, R. and {Honscheid}, K. and {Juneau}, S. and {Kisner}, T. and {Kremin}, A. and {Landriau}, M. and {Liu}, R.~H. and {Louis}, T. and {MacCrann}, N. and {de Macorra}, A. and {Madhavacheril}, M. and {Manera}, M. and {Meisner}, A. and {Miquel}, R. and {Moodley}, K. and {Moustakas}, J. and {Mroczkowski}, T. and {Naess}, S. and {Newman}, J. and {Niemack}, M.~D. and {Niz}, G. and {Page}, L. and {Palanque-Delabrouille}, N. and {Partridge}, B. and {Percival}, W.~J. and {Prada}, F. and {Qu}, F.~J. and {Rossi}, G. and {Sanchez}, E. and {Schlegel}, D. and {Schubnell}, M. and {Sehgal}, N. and {Seo}, H. and {Sif{\'o}n}, C. and {Spergel}, D. and {Sprayberry}, D. and {Staggs}, S. and {Tarl{\'e}}, G. and {Vargas}, C. and {Vavagiakis}, E.~M. and {Weaver}, B.~A. and {Wollack}, E.~J. and {Zhou}, R. and {Zou}, H.},
        title = "{Evidence for large baryonic feedback at low and intermediate redshifts from kinematic Sunyaev-Zel'dovich observations with ACT and DESI photometric galaxies}",
      journal = {arXiv e-prints},
         year = 2024,
        month = jul,
          eid = {arXiv:2407.07152},
        pages = {arXiv:2407.07152},
          doi = {10.48550/arXiv.2407.07152},
archivePrefix = {arXiv},
       eprint = {2407.07152},
 primaryClass = {astro-ph.CO},
       adsurl = {https://ui.adsabs.harvard.edu/abs/2024arXiv240707152H}
}

@ARTICLE{HI4PI2016,
       author = {{HI4PI Collaboration} and {Ben Bekhti}, N. and {Fl{\"o}er}, L. and {Keller}, R. and {Kerp}, J. and {Lenz}, D. and {Winkel}, B. and {Bailin}, J. and {Calabretta}, M.~R. and {Dedes}, L. and {Ford}, H.~A. and {Gibson}, B.~K. and {Haud}, U. and {Janowiecki}, S. and {Kalberla}, P.~M.~W. and {Lockman}, F.~J. and {McClure-Griffiths}, N.~M. and {Murphy}, T. and {Nakanishi}, H. and {Pisano}, D.~J. and {Staveley-Smith}, L.},
        title = "{HI4PI: A full-sky H I survey based on EBHIS and GASS}",
      journal = {\aap},
         year = 2016,
        month = oct,
       volume = {594},
          eid = {A116},
        pages = {A116},
          doi = {10.1051/0004-6361/201629178},
archivePrefix = {arXiv},
       eprint = {1610.06175},
 primaryClass = {astro-ph.GA},
       adsurl = {https://ui.adsabs.harvard.edu/abs/2016A&A...594A.116H}
}

@ARTICLE{Hickox2006,
       author = {{Hickox}, Ryan C. and {Markevitch}, Maxim},
        title = "{Absolute Measurement of the Unresolved Cosmic X-Ray Background in the 0.5-8 keV Band with Chandra}",
      journal = {\apj},
         year = 2006,
        month = jul,
       volume = {645},
       number = {1},
        pages = {95-114},
          doi = {10.1086/504070},
archivePrefix = {arXiv},
       eprint = {astro-ph/0512542},
 primaryClass = {astro-ph},
       adsurl = {https://ui.adsabs.harvard.edu/abs/2006ApJ...645...95H}
}

@ARTICLE{Hopkins2026,
       author = {{Hopkins}, Philip F. and {Silich}, Emily and {Sayers}, Jack and {Ponnada}, Sam B. and {Sands}, Isabel},
        title = "{Observational Implications of Cosmic Ray-Inverse Compton 'Boosted' Cool Cores in Clusters}",
      journal = {arXiv e-prints},
         year = 2026,
        month = jan,
          eid = {arXiv:2601.22229},
        pages = {arXiv:2601.22229},
          doi = {10.48550/arXiv.2601.22229},
archivePrefix = {arXiv},
       eprint = {2601.22229},
 primaryClass = {astro-ph.HE},
       adsurl = {https://ui.adsabs.harvard.edu/abs/2026arXiv260122229H}
}

@ARTICLE{Ilani2024,
       author = {{Ilani}, Gideon and {Hou}, Kuan-Chou and {Keshet}, Uri},
        title = "{Excess cataloged X-ray and radio sources at galaxy-cluster virial shocks}",
      journal = {\jcap},
         year = 2024,
        month = oct,
       volume = {2024},
       number = {10},
          eid = {008},
        pages = {008},
          doi = {10.1088/1475-7516/2024/10/008},
archivePrefix = {arXiv},
       eprint = {2402.16946},
 primaryClass = {astro-ph.HE},
       adsurl = {https://ui.adsabs.harvard.edu/abs/2024JCAP...10..008I}
}

@ARTICLE{Kaastra2017,
       author = {{Kaastra}, J.~S.},
        title = "{On the use of C-stat in testing models for X-ray spectra}",
      journal = {\aap},
         year = 2017,
        month = sep,
       volume = {605},
          eid = {A51},
        pages = {A51},
          doi = {10.1051/0004-6361/201629319},
archivePrefix = {arXiv},
       eprint = {1707.09202},
 primaryClass = {astro-ph.HE},
       adsurl = {https://ui.adsabs.harvard.edu/abs/2017A&A...605A..51K}
}

@article{Kass1995,
  title={Bayes factors},
  author={Kass, Robert E and Raftery, Adrian E},
  journal={Journal of the american statistical association},
  volume={90},
  number={430},
  pages={773--795},
  year={1995},
  publisher={Taylor \& Francis}
}

@ARTICLE{Keshet2020,
       author = {{Keshet}, Uri and {Reiss}, Ido and {Hurier}, Guillaume},
        title = "{Coincident Sunyaev-Zel'dovich and Gamma-Ray Signals from Cluster Virial Shocks}",
      journal = {\apj},
         year = 2020,
        month = may,
       volume = {895},
       number = {1},
          eid = {72},
        pages = {72},
          doi = {10.3847/1538-4357/ab8c49},
archivePrefix = {arXiv},
       eprint = {1801.01494},
 primaryClass = {astro-ph.CO},
       adsurl = {https://ui.adsabs.harvard.edu/abs/2020ApJ...895...72K}
}

@ARTICLE{Kleinebreil2025,
       author = {{Kleinebreil}, F. and {Grandis}, S. and {Schrabback}, T. and {Ghirardini}, V. and {Chiu}, I. and {Liu}, A. and {Kluge}, M. and {Reiprich}, T.~H. and {Artis}, E. and {Bahar}, Y.~E. and {Balzer}, F. and {Bulbul}, E. and {Clerc}, N. and {Comparat}, J. and {Garrel}, C. and {Gruen}, D. and {Li}, X. and {Miyatake}, H. and {Miyazaki}, S. and {Ramos-Ceja}, M.~E. and {Sanders}, J. and {Seppi}, R. and {Okabe}, N. and {Zhang}, X.},
        title = "{The SRG/eROSITA All-Sky Survey: Weak lensing of eRASS1 galaxy clusters in KiDS-1000 and consistency checks with DES Y3 and HSC-Y3}",
      journal = {\aap},
         year = 2025,
        month = mar,
       volume = {695},
          eid = {A216},
        pages = {A216},
          doi = {10.1051/0004-6361/202449599},
archivePrefix = {arXiv},
       eprint = {2402.08456},
 primaryClass = {astro-ph.CO},
       adsurl = {https://ui.adsabs.harvard.edu/abs/2025A&A...695A.216K}
}

@ARTICLE{Kluge2024,
       author = {{Kluge}, M. and {Comparat}, J. and {Liu}, A. and {Balzer}, F. and {Bulbul}, E. and {Ider Chitham}, J. and {Ghirardini}, V. and {Garrel}, C. and {Bahar}, Y.~E. and {Artis}, E. and {Bender}, R. and {Clerc}, N. and {Dwelly}, T. and {Fabricius}, M.~H. and {Grandis}, S. and {Hern{\'a}ndez-Lang}, D. and {Hill}, G.~J. and {Joshi}, J. and {Lamer}, G. and {Merloni}, A. and {Nandra}, K. and {Pacaud}, F. and {Predehl}, P. and {Ramos-Ceja}, M.~E. and {Reiprich}, T.~H. and {Salvato}, M. and {Sanders}, J.~S. and {Schrabback}, T. and {Seppi}, R. and {Zelmer}, S. and {Zenteno}, A. and {Zhang}, X.},
        title = "{The First SRG/eROSITA All-Sky Survey: Optical Identification and Properties of Galaxy Clusters and Groups in the Western Galactic Hemisphere}",
      journal = {arXiv e-prints},
         year = 2024,
        month = feb,
          eid = {arXiv:2402.08453},
        pages = {arXiv:2402.08453},
          doi = {10.48550/arXiv.2402.08453},
archivePrefix = {arXiv},
       eprint = {2402.08453},
 primaryClass = {astro-ph.CO},
       adsurl = {https://ui.adsabs.harvard.edu/abs/2024arXiv240208453K}
}

@ARTICLE{Kuntz2015,
       author = {{Kuntz}, K.~D. and {Collado-Vega}, Y.~M. and {Collier}, M.~R. and {Connor}, H.~K. and {Cravens}, T.~E. and {Koutroumpa}, D. and {Porter}, F.~S. and {Robertson}, I.~P. and {Sibeck}, D.~G. and {Snowden}, S.~L. and {Thomas}, N.~E. and {Walsh}, B.~M.},
        title = "{The Solar Wind Charge-exchange Production Factor for Hydrogen}",
      journal = {\apj},
         year = 2015,
        month = aug,
       volume = {808},
       number = {2},
          eid = {143},
        pages = {143},
          doi = {10.1088/0004-637X/808/2/143},
archivePrefix = {arXiv},
       eprint = {1503.04756},
 primaryClass = {astro-ph.HE},
       adsurl = {https://ui.adsabs.harvard.edu/abs/2015ApJ...808..143K}
}

@ARTICLE{Liu2022,
       author = {{Liu}, A. and {Bulbul}, E. and {Ghirardini}, V. and {Liu}, T. and {Klein}, M. and {Clerc}, N. and {{\"O}zsoy}, Y. and {Ramos-Ceja}, M.~E. and {Pacaud}, F. and {Comparat}, J. and {Okabe}, N. and {Bahar}, Y.~E. and {Biffi}, V. and {Brunner}, H. and {Br{\"u}ggen}, M. and {Buchner}, J. and {Ider Chitham}, J. and {Chiu}, I. and {Dolag}, K. and {Gatuzz}, E. and {Gonzalez}, J. and {Hoang}, D.~N. and {Lamer}, G. and {Merloni}, A. and {Nandra}, K. and {Oguri}, M. and {Ota}, N. and {Predehl}, P. and {Reiprich}, T.~H. and {Salvato}, M. and {Schrabback}, T. and {Sanders}, J.~S. and {Seppi}, R. and {Thibaud}, Q.},
        title = "{The eROSITA Final Equatorial-Depth Survey (eFEDS). Catalog of galaxy clusters and groups}",
      journal = {\aap},
         year = 2022,
        month = may,
       volume = {661},
          eid = {A2},
        pages = {A2},
          doi = {10.1051/0004-6361/202141120},
archivePrefix = {arXiv},
       eprint = {2106.14518},
 primaryClass = {astro-ph.CO},
       adsurl = {https://ui.adsabs.harvard.edu/abs/2022A&A...661A...2L}
}

@ARTICLE{Liu2023,
       author = {{Liu}, A. and {Bulbul}, E. and {Ramos-Ceja}, M.~E. and {Sanders}, J.~S. and {Ghirardini}, V. and {Bahar}, Y.~E. and {Yeung}, M. and {Gatuzz}, E. and {Freyberg}, M. and {Garrel}, C. and {Zhang}, X. and {Merloni}, A. and {Nandra}, K.},
        title = "{X-ray analysis of JWST's first galaxy cluster lens SMACS J0723.3‒7327}",
      journal = {\aap},
         year = 2023,
        month = feb,
       volume = {670},
          eid = {A96},
        pages = {A96},
          doi = {10.1051/0004-6361/202245118},
archivePrefix = {arXiv},
       eprint = {2210.00633},
 primaryClass = {astro-ph.CO},
       adsurl = {https://ui.adsabs.harvard.edu/abs/2023A&A...670A..96L}
}

@ARTICLE{Lodders2003,
       author = {{Lodders}, Katharina},
        title = "{Solar System Abundances and Condensation Temperatures of the Elements}",
      journal = {\apj},
         year = 2003,
        month = jul,
       volume = {591},
       number = {2},
        pages = {1220-1247},
          doi = {10.1086/375492},
       adsurl = {https://ui.adsabs.harvard.edu/abs/2003ApJ...591.1220L}
}

@ARTICLE{Malavasi2020,
       author = {{Malavasi}, Nicola and {Aghanim}, Nabila and {Douspis}, Marian and {Tanimura}, Hideki and {Bonjean}, Victor},
        title = "{Characterising filaments in the SDSS volume from the galaxy distribution}",
      journal = {\aap},
         year = 2020,
        month = oct,
       volume = {642},
          eid = {A19},
        pages = {A19},
          doi = {10.1051/0004-6361/202037647},
archivePrefix = {arXiv},
       eprint = {2002.01486},
 primaryClass = {astro-ph.CO},
       adsurl = {https://ui.adsabs.harvard.edu/abs/2020A&A...642A..19M}
}

@ARTICLE{Marini2025,
       author = {{Marini}, I. and {Popesso}, P. and {Dolag}, K. and {Bravo}, M. and {Robotham}, A. and {Tempel}, E. and {Li}, Q. and {Yang}, X. and {Csizi}, B. and {Behroozi}, P. and {Biffi}, V. and {Biviano}, A. and {Lamer}, G. and {Malavasi}, N. and {Mazengo}, D. and {Toptun}, V.},
        title = "{Detecting clusters and groups of galaxies populating the local Universe in large optical spectroscopic surveys}",
      journal = {\aap},
         year = 2025,
        month = feb,
       volume = {694},
          eid = {A207},
        pages = {A207},
          doi = {10.1051/0004-6361/202452028},
archivePrefix = {arXiv},
       eprint = {2411.16455},
 primaryClass = {astro-ph.GA},
       adsurl = {https://ui.adsabs.harvard.edu/abs/2025A&A...694A.207M}
}

@ARTICLE{McNamara2007,
       author = {{McNamara}, B.~R. and {Nulsen}, P.~E.~J.},
        title = "{Heating Hot Atmospheres with Active Galactic Nuclei}",
      journal = {\araa},
         year = 2007,
        month = sep,
       volume = {45},
       number = {1},
        pages = {117-175},
          doi = {10.1146/annurev.astro.45.051806.110625},
archivePrefix = {arXiv},
       eprint = {0709.2152},
 primaryClass = {astro-ph},
       adsurl = {https://ui.adsabs.harvard.edu/abs/2007ARA&A..45..117M}
}

@ARTICLE{Merloni2024,
       author = {{Merloni}, A. and {Lamer}, G. and {Liu}, T. and {Ramos-Ceja}, M.~E. and {Brunner}, H. and {Bulbul}, E. and {Dennerl}, K. and {Doroshenko}, V. and {Freyberg}, M.~J. and {Friedrich}, S. and {Gatuzz}, E. and {Georgakakis}, A. and {Haberl}, F. and {Igo}, Z. and {Kreykenbohm}, I. and {Liu}, A. and {Maitra}, C. and {Malyali}, A. and {Mayer}, M.~G.~F. and {Nandra}, K. and {Predehl}, P. and {Robrade}, J. and {Salvato}, M. and {Sanders}, J.~S. and {Stewart}, I. and {Tub{\'\i}n-Arenas}, D. and {Weber}, P. and {Wilms}, J. and {Arcodia}, R. and {Artis}, E. and {Aschersleben}, J. and {Avakyan}, A. and {Aydar}, C. and {Bahar}, Y.~E. and {Balzer}, F. and {Becker}, W. and {Berger}, K. and {Boller}, T. and {Bornemann}, W. and {Br{\"u}ggen}, M. and {Brusa}, M. and {Buchner}, J. and {Burwitz}, V. and {Camilloni}, F. and {Clerc}, N. and {Comparat}, J. and {Coutinho}, D. and {Czesla}, S. and {Dannhauer}, S.~M. and {Dauner}, L. and {Dauser}, T. and {Dietl}, J. and {Dolag}, K. and {Dwelly}, T. and {Egg}, K. and {Ehl}, E. and {Freund}, S. and {Friedrich}, P. and {Gaida}, R. and {Garrel}, C. and {Ghirardini}, V. and {Gokus}, A. and {Gr{\"u}nwald}, G. and {Grandis}, S. and {Grotova}, I. and {Gruen}, D. and {Gueguen}, A. and {H{\"a}mmerich}, S. and {Hamaus}, N. and {Hasinger}, G. and {Haubner}, K. and {Homan}, D. and {Ider Chitham}, J. and {Joseph}, W.~M. and {Joyce}, A. and {K{\"o}nig}, O. and {Kaltenbrunner}, D.~M. and {Khokhriakova}, A. and {Kink}, W. and {Kirsch}, C. and {Kluge}, M. and {Knies}, J. and {Krippendorf}, S. and {Krumpe}, M. and {Kurpas}, J. and {Li}, P. and {Liu}, Z. and {Locatelli}, N. and {Lorenz}, M. and {M{\"u}ller}, S. and {Magaudda}, E. and {Mannes}, C. and {McCall}, H. and {Meidinger}, N. and {Michailidis}, M. and {Migkas}, K. and {Mu{\~n}oz-Giraldo}, D. and {Musiimenta}, B. and {Nguyen-Dang}, N.~T. and {Ni}, Q. and {Olechowska}, A. and {Ota}, N. and {Pacaud}, F. and {Pasini}, T. and {Perinati}, E. and {Pires}, A.~M. and {Pommranz}, C. and {Ponti}, G. and {Poppenhaeger}, K. and {P{\"u}hlhofer}, G. and {Rau}, A. and {Reh}, M. and {Reiprich}, T.~H. and {Roster}, W. and {Saeedi}, S. and {Santangelo}, A. and {Sasaki}, M. and {Schmitt}, J. and {Schneider}, P.~C. and {Schrabback}, T. and {Schuster}, N. and {Schwope}, A. and {Seppi}, R. and {Serim}, M.~M. and {Shreeram}, S. and {Sokolova-Lapa}, E. and {Starck}, H. and {Stelzer}, B. and {Stierhof}, J. and {Suleimanov}, V. and {Tenzer}, C. and {Traulsen}, I. and {Tr{\"u}mper}, J. and {Tsuge}, K. and {Urrutia}, T. and {Veronica}, A. and {Waddell}, S.~G.~H. and {Willer}, R. and {Wolf}, J. and {Yeung}, M.~C.~H. and {Zainab}, A. and {Zangrandi}, F. and {Zhang}, X. and {Zhang}, Y. and {Zheng}, X.},
        title = "{The SRG/eROSITA all-sky survey. First X-ray catalogues and data release of the western Galactic hemisphere}",
      journal = {\aap},
         year = 2024,
        month = feb,
       volume = {682},
          eid = {A34},
        pages = {A34},
          doi = {10.1051/0004-6361/202347165},
archivePrefix = {arXiv},
       eprint = {2401.17274},
 primaryClass = {astro-ph.HE},
       adsurl = {https://ui.adsabs.harvard.edu/abs/2024A&A...682A..34M}
}

@ARTICLE{Mernier2016,
       author = {{Mernier}, F. and {de Plaa}, J. and {Pinto}, C. and {Kaastra}, J.~S. and {Kosec}, P. and {Zhang}, Y.-Y. and {Mao}, J. and {Werner}, N. and {Pols}, O.~R. and {Vink}, J.},
        title = "{Origin of central abundances in the hot intra-cluster medium. II. Chemical enrichment and supernova yield models}",
      journal = {\aap},
         year = 2016,
        month = nov,
       volume = {595},
          eid = {A126},
        pages = {A126},
          doi = {10.1051/0004-6361/201628765},
archivePrefix = {arXiv},
       eprint = {1608.03888},
 primaryClass = {astro-ph.GA},
       adsurl = {https://ui.adsabs.harvard.edu/abs/2016A&A...595A.126M}
}

@ARTICLE{Mernier2022,
       author = {{Mernier}, F. and {Werner}, N. and {Su}, Y. and {Pinto}, C. and {Grossov{\'a}}, R. and {Simionescu}, A. and {Iodice}, E. and {Sarzi}, M. and {G{\"o}rgei}, A.},
        title = "{The cycle of metals in the infalling elliptical galaxy NGC 1404}",
      journal = {\mnras},
         year = 2022,
        month = apr,
       volume = {511},
       number = {3},
        pages = {3159-3178},
          doi = {10.1093/mnras/stac253},
archivePrefix = {arXiv},
       eprint = {2201.05161},
 primaryClass = {astro-ph.GA},
       adsurl = {https://ui.adsabs.harvard.edu/abs/2022MNRAS.511.3159M}
}

@ARTICLE{Mernier2023,
       author = {{Mernier}, F. and {Werner}, N. and {Bagchi}, J. and {Gendron-Marsolais}, M.-L. and {Gopal-Krishna} and {Guainazzi}, M. and {Richard-Laferri{\`e}re}, A. and {Shimwell}, T.~W. and {Simionescu}, A.},
        title = "{Discovery of inverse-Compton X-ray emission and estimate of the volume-averaged magnetic field in a galaxy group}",
      journal = {\mnras},
         year = 2023,
        month = oct,
       volume = {524},
       number = {4},
        pages = {4939-4949},
          doi = {10.1093/mnras/stad2093},
archivePrefix = {arXiv},
       eprint = {2207.10092},
 primaryClass = {astro-ph.GA},
       adsurl = {https://ui.adsabs.harvard.edu/abs/2023MNRAS.524.4939M}
}

@ARTICLE{Migkas2024,
       author = {{Migkas}, K. and {Kox}, D. and {Schellenberger}, G. and {Veronica}, A. and {Pacaud}, F. and {Reiprich}, T.~H. and {Bahar}, Y.~E. and {Balzer}, F. and {Bulbul}, E. and {Comparat}, J. and {Dennerl}, K. and {Freyberg}, M. and {Garrel}, C. and {Ghirardini}, V. and {Grandis}, S. and {Kluge}, M. and {Liu}, A. and {Ramos-Ceja}, M.~E. and {Sanders}, J. and {Zhang}, X.},
        title = "{The SRG/eROSITA All-Sky Survey. SRG/eROSITA cross-calibration with Chandra and XMM-Newton using galaxy cluster gas temperatures}",
      journal = {\aap},
         year = 2024,
        month = aug,
       volume = {688},
          eid = {A107},
        pages = {A107},
          doi = {10.1051/0004-6361/202349006},
archivePrefix = {arXiv},
       eprint = {2401.17297},
 primaryClass = {astro-ph.CO},
       adsurl = {https://ui.adsabs.harvard.edu/abs/2024A&A...688A.107M}
}

@ARTICLE{Molnar2009,
       author = {{Molnar}, Sandor M. and {Hearn}, Nathan and {Haiman}, Zolt{\'a}n and {Bryan}, Greg and {Evrard}, August E. and {Lake}, George},
        title = "{Accretion Shocks in Clusters of Galaxies and Their SZ Signature from Cosmological Simulations}",
      journal = {\apj},
         year = 2009,
        month = may,
       volume = {696},
       number = {2},
        pages = {1640-1656},
          doi = {10.1088/0004-637X/696/2/1640},
archivePrefix = {arXiv},
       eprint = {0902.3323},
 primaryClass = {astro-ph.CO},
       adsurl = {https://ui.adsabs.harvard.edu/abs/2009ApJ...696.1640M}
}

@ARTICLE{Moretti2012,
       author = {{Moretti}, A. and {Vattakunnel}, S. and {Tozzi}, P. and {Salvaterra}, R. and {Severgnini}, P. and {Fugazza}, D. and {Haardt}, F. and {Gilli}, R.},
        title = "{Spectrum of the unresolved cosmic X-ray background: what is unresolved 50 years after its discovery}",
      journal = {\aap},
         year = 2012,
        month = dec,
       volume = {548},
          eid = {A87},
        pages = {A87},
          doi = {10.1051/0004-6361/201219921},
archivePrefix = {arXiv},
       eprint = {1210.6377},
 primaryClass = {astro-ph.CO},
       adsurl = {https://ui.adsabs.harvard.edu/abs/2012A&A...548A..87M}
}

@ARTICLE{Nandra2013,
       author = {{Nandra}, Kirpal and {Barret}, Didier and {Barcons}, Xavier and {Fabian}, Andy and {den Herder}, Jan-Willem and {Piro}, Luigi and {Watson}, Mike and {Adami}, Christophe and {Aird}, James and {Afonso}, Jose Manuel and {Alexander}, Dave and {Argiroffi}, Costanza and {Amati}, Lorenzo and {Arnaud}, Monique and {Atteia}, Jean-Luc and {Audard}, Marc and {Badenes}, Carles and {Ballet}, Jean and {Ballo}, Lucia and {Bamba}, Aya and {Bhardwaj}, Anil and {Stefano Battistelli}, Elia and {Becker}, Werner and {De Becker}, Micha{\"e}l and {Behar}, Ehud and {Bianchi}, Stefano and {Biffi}, Veronica and {B{\^\i}rzan}, Laura and {Bocchino}, Fabrizio and {Bogdanov}, Slavko and {Boirin}, Laurence and {Boller}, Thomas and {Borgani}, Stefano and {Borm}, Katharina and {Bouch{\'e}}, Nicolas and {Bourdin}, Herv{\'e} and {Bower}, Richard and {Braito}, Valentina and {Branchini}, Enzo and {Branduardi-Raymont}, Graziella and {Bregman}, Joel and {Brenneman}, Laura and {Brightman}, Murray and {Br{\"u}ggen}, Marcus and {Buchner}, Johannes and {Bulbul}, Esra and {Brusa}, Marcella and {Bursa}, Michal and {Caccianiga}, Alessandro and {Cackett}, Ed and {Campana}, Sergio and {Cappelluti}, Nico and {Cappi}, Massimo and {Carrera}, Francisco and {Ceballos}, Maite and {Christensen}, Finn and {Chu}, You-Hua and {Churazov}, Eugene and {Clerc}, Nicolas and {Corbel}, Stephane and {Corral}, Amalia and {Comastri}, Andrea and {Costantini}, Elisa and {Croston}, Judith and {Dadina}, Mauro and {D'Ai}, Antonino and {Decourchelle}, Anne and {Della Ceca}, Roberto and {Dennerl}, Konrad and {Dolag}, Klaus and {Done}, Chris and {Dovciak}, Michal and {Drake}, Jeremy and {Eckert}, Dominique and {Edge}, Alastair and {Ettori}, Stefano and {Ezoe}, Yuichiro and {Feigelson}, Eric and {Fender}, Rob and {Feruglio}, Chiara and {Finoguenov}, Alexis and {Fiore}, Fabrizio and {Galeazzi}, Massimiliano and {Gallagher}, Sarah and {Gandhi}, Poshak and {Gaspari}, Massimo and {Gastaldello}, Fabio and {Georgakakis}, Antonis and {Georgantopoulos}, Ioannis and {Gilfanov}, Marat and {Gitti}, Myriam and {Gladstone}, Randy and {Goosmann}, Rene and {Gosset}, Eric and {Grosso}, Nicolas and {Guedel}, Manuel and {Guerrero}, Martin and {Haberl}, Frank and {Hardcastle}, Martin and {Heinz}, Sebastian and {Alonso Herrero}, Almudena and {Herv{\'e}}, Anthony and {Holmstrom}, Mats and {Iwasawa}, Kazushi and {Jonker}, Peter and {Kaastra}, Jelle and {Kara}, Erin and {Karas}, Vladimir and {Kastner}, Joel and {King}, Andrew and {Kosenko}, Daria and {Koutroumpa}, Dimita and {Kraft}, Ralph and {Kreykenbohm}, Ingo and {Lallement}, Rosine and {Lanzuisi}, Giorgio and {Lee}, J. and {Lemoine-Goumard}, Marianne and {Lobban}, Andrew and {Lodato}, Giuseppe and {Lovisari}, Lorenzo and {Lotti}, Simone and {McCharthy}, Ian and {McNamara}, Brian and {Maggio}, Antonio and {Maiolino}, Roberto and {De Marco}, Barbara and {de Martino}, Domitilla and {Mateos}, Silvia and {Matt}, Giorgio and {Maughan}, Ben and {Mazzotta}, Pasquale and {Mendez}, Mariano and {Merloni}, Andrea and {Micela}, Giuseppina and {Miceli}, Marco and {Mignani}, Robert and {Miller}, Jon and {Miniutti}, Giovanni and {Molendi}, Silvano and {Montez}, Rodolfo and {Moretti}, Alberto and {Motch}, Christian and {Naz{\'e}}, Ya{\"e}l and {Nevalainen}, Jukka and {Nicastro}, Fabrizio and {Nulsen}, Paul and {Ohashi}, Takaya and {O'Brien}, Paul and {Osborne}, Julian and {Oskinova}, Lida and {Pacaud}, Florian and {Paerels}, Frederik and {Page}, Mat and {Papadakis}, Iossif and {Pareschi}, Giovanni and {Petre}, Robert and {Petrucci}, Pierre-Olivier and {Piconcelli}, Enrico and {Pillitteri}, Ignazio and {Pinto}, C. and {de Plaa}, Jelle and {Pointecouteau}, Etienne and {Ponman}, Trevor and {Ponti}, Gabriele and {Porquet}, Delphine and {Pounds}, Ken and {Pratt}, Gabriel and {Predehl}, Peter and {Proga}, Daniel and {Psaltis}, Dimitrios and {Rafferty}, David and {Ramos-Ceja}, Miriam and {Ranalli}, Piero and {Rasia}, Elena and {Rau}, Arne and {Rauw}, Gregor and {Rea}, Nanda and {Read}, Andy and {Reeves}, James and {Reiprich}, Thomas and {Renaud}, Matthieu and {Reynolds}, Chris and {Risaliti}, Guido and {Rodriguez}, Jerome and {Rodriguez Hidalgo}, Paola and {Roncarelli}, Mauro and {Rosario}, David and {Rossetti}, Mariachiara and {Rozanska}, Agata and {Rovilos}, Emmanouil and {Salvaterra}, Ruben and {Salvato}, Mara and {Di Salvo}, Tiziana and {Sanders}, Jeremy and {Sanz-Forcada}, Jorge and {Schawinski}, Kevin and {Schaye}, Joop and {Schwope}, Axel and {Sciortino}, Salvatore},
        title = "{The Hot and Energetic Universe: A White Paper presenting the science theme motivating the Athena+ mission}",
      journal = {arXiv e-prints},
         year = 2013,
        month = jun,
          eid = {arXiv:1306.2307},
        pages = {arXiv:1306.2307},
          doi = {10.48550/arXiv.1306.2307},
archivePrefix = {arXiv},
       eprint = {1306.2307},
 primaryClass = {astro-ph.HE},
       adsurl = {https://ui.adsabs.harvard.edu/abs/2013arXiv1306.2307N}
}

@ARTICLE{Oei2024,
       author = {{Oei}, Martijn S.~S.~L. and {Hardcastle}, Martin J. and {Timmerman}, Roland and {Gast}, Aivin R.~D.~J.~G.~I.~B. and {Botteon}, Andrea and {Rodriguez}, Antonio C. and {Stern}, Daniel and {Calistro Rivera}, Gabriela and {van Weeren}, Reinout J. and {R{\"o}ttgering}, Huub J.~A. and {Intema}, Huib T. and {de Gasperin}, Francesco and {Djorgovski}, S.~G.},
        title = "{Black hole jets on the scale of the cosmic web}",
      journal = {\nat},
         year = 2024,
        month = sep,
       volume = {633},
       number = {8030},
        pages = {537-541},
          doi = {10.1038/s41586-024-07879-y},
archivePrefix = {arXiv},
       eprint = {2411.08630},
 primaryClass = {astro-ph.HE},
       adsurl = {https://ui.adsabs.harvard.edu/abs/2024Natur.633..537O}
}

@ARTICLE{Okabe2025,
       author = {{Okabe}, Nobuhiro and {Reiprich}, Thomas and {Grandis}, Sebastian and {Chiu}, I-Non and {Oguri}, Masamune and {Umetsu}, Keiichi and {Bulbul}, Esra and {Bahar}, Emre and {Balzer}, Fabian and {Clerc}, Nicolas and {Comparat}, Johan and {Ghirardini}, Vittorio and {Kleinebreil}, Florian and {Kluge}, Matthias and {Liu}, Ang and {Lin}, Yen-Ting and {Monteiro-Oliveira}, Rog{\'e}rio and {Pacaud}, Florian and {Ramos Ceja}, Miriam and {Sanders}, Jeremy and {Schrabback}, Tim and {Seppi}, Riccardo and {Sommer}, Martin and {Zhang}, Xiaoyuan},
        title = "{The SRG/eROSITA All-Sky Survey : Subaru/HSC-SSP weak-lensing mass measurements for the eRASS1 Galaxy Clusters}",
      journal = {arXiv e-prints},
         year = 2025,
        month = mar,
          eid = {arXiv:2503.09952},
        pages = {arXiv:2503.09952},
          doi = {10.48550/arXiv.2503.09952},
archivePrefix = {arXiv},
       eprint = {2503.09952},
 primaryClass = {astro-ph.CO},
       adsurl = {https://ui.adsabs.harvard.edu/abs/2025arXiv250309952O}
}

@ARTICLE{OSullivan2011b,
       author = {{O'Sullivan}, E. and {Giacintucci}, S. and {David}, L.~P. and {Gitti}, M. and {Vrtilek}, J.~M. and {Raychaudhury}, S. and {Ponman}, T.~J.},
        title = "{Heating the Hot Atmospheres of Galaxy Groups and Clusters with Cavities: The Relationship between Jet Power and Low-frequency Radio Emission}",
      journal = {\apj},
         year = 2011,
        month = jul,
       volume = {735},
       number = {1},
          eid = {11},
        pages = {11},
          doi = {10.1088/0004-637X/735/1/11},
archivePrefix = {arXiv},
       eprint = {1104.2411},
 primaryClass = {astro-ph.CO},
       adsurl = {https://ui.adsabs.harvard.edu/abs/2011ApJ...735...11O}
}

@ARTICLE{Petrosian2001,
       author = {{Petrosian}, Vah{\'e}},
        title = "{On the Nonthermal Emission and Acceleration of Electrons in Coma and Other Clusters of Galaxies}",
      journal = {\apj},
         year = 2001,
        month = aug,
       volume = {557},
       number = {2},
        pages = {560-572},
          doi = {10.1086/321557},
archivePrefix = {arXiv},
       eprint = {astro-ph/0101145},
 primaryClass = {astro-ph},
       adsurl = {https://ui.adsabs.harvard.edu/abs/2001ApJ...557..560P}
}

@ARTICLE{Planck2020,
       author = {{Planck Collaboration} and {Aghanim}, N. and {Akrami}, Y. and {Ashdown}, M. and {Aumont}, J. and {Baccigalupi}, C. and {Ballardini}, M. and {Banday}, A.~J. and {Barreiro}, R.~B. and {Bartolo}, N. and {Basak}, S. and {Battye}, R. and {Benabed}, K. and {Bernard}, J.-P. and {Bersanelli}, M. and {Bielewicz}, P. and {Bock}, J.~J. and {Bond}, J.~R. and {Borrill}, J. and {Bouchet}, F.~R. and {Boulanger}, F. and {Bucher}, M. and {Burigana}, C. and {Butler}, R.~C. and {Calabrese}, E. and {Cardoso}, J.-F. and {Carron}, J. and {Challinor}, A. and {Chiang}, H.~C. and {Chluba}, J. and {Colombo}, L.~P.~L. and {Combet}, C. and {Contreras}, D. and {Crill}, B.~P. and {Cuttaia}, F. and {de Bernardis}, P. and {de Zotti}, G. and {Delabrouille}, J. and {Delouis}, J.-M. and {Di Valentino}, E. and {Diego}, J.~M. and {Dor{\'e}}, O. and {Douspis}, M. and {Ducout}, A. and {Dupac}, X. and {Dusini}, S. and {Efstathiou}, G. and {Elsner}, F. and {En{\ss}lin}, T.~A. and {Eriksen}, H.~K. and {Fantaye}, Y. and {Farhang}, M. and {Fergusson}, J. and {Fernandez-Cobos}, R. and {Finelli}, F. and {Forastieri}, F. and {Frailis}, M. and {Fraisse}, A.~A. and {Franceschi}, E. and {Frolov}, A. and {Galeotta}, S. and {Galli}, S. and {Ganga}, K. and {G{\'e}nova-Santos}, R.~T. and {Gerbino}, M. and {Ghosh}, T. and {Gonz{\'a}lez-Nuevo}, J. and {G{\'o}rski}, K.~M. and {Gratton}, S. and {Gruppuso}, A. and {Gudmundsson}, J.~E. and {Hamann}, J. and {Handley}, W. and {Hansen}, F.~K. and {Herranz}, D. and {Hildebrandt}, S.~R. and {Hivon}, E. and {Huang}, Z. and {Jaffe}, A.~H. and {Jones}, W.~C. and {Karakci}, A. and {Keih{\"a}nen}, E. and {Keskitalo}, R. and {Kiiveri}, K. and {Kim}, J. and {Kisner}, T.~S. and {Knox}, L. and {Krachmalnicoff}, N. and {Kunz}, M. and {Kurki-Suonio}, H. and {Lagache}, G. and {Lamarre}, J.-M. and {Lasenby}, A. and {Lattanzi}, M. and {Lawrence}, C.~R. and {Le Jeune}, M. and {Lemos}, P. and {Lesgourgues}, J. and {Levrier}, F. and {Lewis}, A. and {Liguori}, M. and {Lilje}, P.~B. and {Lilley}, M. and {Lindholm}, V. and {L{\'o}pez-Caniego}, M. and {Lubin}, P.~M. and {Ma}, Y.-Z. and {Mac{\'\i}as-P{\'e}rez}, J.~F. and {Maggio}, G. and {Maino}, D. and {Mandolesi}, N. and {Mangilli}, A. and {Marcos-Caballero}, A. and {Maris}, M. and {Martin}, P.~G. and {Martinelli}, M. and {Mart{\'\i}nez-Gonz{\'a}lez}, E. and {Matarrese}, S. and {Mauri}, N. and {McEwen}, J.~D. and {Meinhold}, P.~R. and {Melchiorri}, A. and {Mennella}, A. and {Migliaccio}, M. and {Millea}, M. and {Mitra}, S. and {Miville-Desch{\^e}nes}, M.-A. and {Molinari}, D. and {Montier}, L. and {Morgante}, G. and {Moss}, A. and {Natoli}, P. and {N{\o}rgaard-Nielsen}, H.~U. and {Pagano}, L. and {Paoletti}, D. and {Partridge}, B. and {Patanchon}, G. and {Peiris}, H.~V. and {Perrotta}, F. and {Pettorino}, V. and {Piacentini}, F. and {Polastri}, L. and {Polenta}, G. and {Puget}, J.-L. and {Rachen}, J.~P. and {Reinecke}, M. and {Remazeilles}, M. and {Renzi}, A. and {Rocha}, G. and {Rosset}, C. and {Roudier}, G. and {Rubi{\~n}o-Mart{\'\i}n}, J.~A. and {Ruiz-Granados}, B. and {Salvati}, L. and {Sandri}, M. and {Savelainen}, M. and {Scott}, D. and {Shellard}, E.~P.~S. and {Sirignano}, C. and {Sirri}, G. and {Spencer}, L.~D. and {Sunyaev}, R. and {Suur-Uski}, A.-S. and {Tauber}, J.~A. and {Tavagnacco}, D. and {Tenti}, M. and {Toffolatti}, L. and {Tomasi}, M. and {Trombetti}, T. and {Valenziano}, L. and {Valiviita}, J. and {Van Tent}, B. and {Vibert}, L. and {Vielva}, P. and {Villa}, F. and {Vittorio}, N. and {Wandelt}, B.~D. and {Wehus}, I.~K. and {White}, M. and {White}, S.~D.~M. and {Zacchei}, A. and {Zonca}, A.},
        title = "{Planck 2018 results. VI. Cosmological parameters}",
      journal = {\aap},
         year = 2020,
        month = sep,
       volume = {641},
          eid = {A6},
        pages = {A6},
          doi = {10.1051/0004-6361/201833910},
archivePrefix = {arXiv},
       eprint = {1807.06209},
 primaryClass = {astro-ph.CO},
       adsurl = {https://ui.adsabs.harvard.edu/abs/2020A&A...641A...6P}
}

@article{Predehl2021,
	adsurl = {https://ui.adsabs.harvard.edu/abs/2021A&A...647A...1P},
	archiveprefix = {arXiv},
	author = {{Predehl}, P. and {Andritschke}, R. and {Arefiev}, V. and {Babyshkin}, V. and {Batanov}, O. and {Becker}, W. and {B{\"o}hringer}, H. and {Bogomolov}, A. and {Boller}, T. and {Borm}, K. and {Bornemann}, W. and {Br{\"a}uninger}, H. and {Br{\"u}ggen}, M. and {Brunner}, H. and {Brusa}, M. and {Bulbul}, E. and {Buntov}, M. and {Burwitz}, V. and {Burkert}, W. and {Clerc}, N. and {Churazov}, E. and {Coutinho}, D. and {Dauser}, T. and {Dennerl}, K. and {Doroshenko}, V. and {Eder}, J. and {Emberger}, V. and {Eraerds}, T. and {Finoguenov}, A. and {Freyberg}, M. and {Friedrich}, P. and {Friedrich}, S. and {F{\"u}rmetz}, M. and {Georgakakis}, A. and {Gilfanov}, M. and {Granato}, S. and {Grossberger}, C. and {Gueguen}, A. and {Gureev}, P. and {Haberl}, F. and {H{\"a}lker}, O. and {Hartner}, G. and {Hasinger}, G. and {Huber}, H. and {Ji}, L. and {Kienlin}, A. v. and {Kink}, W. and {Korotkov}, F. and {Kreykenbohm}, I. and {Lamer}, G. and {Lomakin}, I. and {Lapshov}, I. and {Liu}, T. and {Maitra}, C. and {Meidinger}, N. and {Menz}, B. and {Merloni}, A. and {Mernik}, T. and {Mican}, B. and {Mohr}, J. and {M{\"u}ller}, S. and {Nandra}, K. and {Nazarov}, V. and {Pacaud}, F. and {Pavlinsky}, M. and {Perinati}, E. and {Pfeffermann}, E. and {Pietschner}, D. and {Ramos-Ceja}, M.~E. and {Rau}, A. and {Reiffers}, J. and {Reiprich}, T.~H. and {Robrade}, J. and {Salvato}, M. and {Sanders}, J. and {Santangelo}, A. and {Sasaki}, M. and {Scheuerle}, H. and {Schmid}, C. and {Schmitt}, J. and {Schwope}, A. and {Shirshakov}, A. and {Steinmetz}, M. and {Stewart}, I. and {Str{\"u}der}, L. and {Sunyaev}, R. and {Tenzer}, C. and {Tiedemann}, L. and {Tr{\"u}mper}, J. and {Voron}, V. and {Weber}, P. and {Wilms}, J. and {Yaroshenko}, V.},
	doi = {10.1051/0004-6361/202039313},
	eid = {A1},
	eprint = {2010.03477},
	journal = {\aap},
	month = mar,
	pages = {A1},
	primaryclass = {astro-ph.HE},
	title = {{The eROSITA X-ray telescope on SRG}},
	volume = {647},
	year = 2021}

@ARTICLE{Popesso2024,
       author = {{Popesso}, P. and {Biviano}, A. and {Marini}, I. and {Dolag}, K. and {Vladutescu-Zopp}, S. and {Csizi}, B. and {Biffi}, V. and {Lamer}, G. and {Robothan}, A. and {Bravo}, M. and {Lovisari}, L. and {Ettori}, S. and {Angelinelli}, M. and {Driver}, S. and {Toptun}, V. and {Dev}, A. and {Mazengo}, D. and {Merloni}, A. and {Comparat}, J. and {Ponti}, G. and {Mroczkowski}, T. and {Bulbul}, E. and {Grandis}, S. and {Bahar}, E.},
        title = "{The hot gas mass fraction in halos. From Milky Way-like groups to massive clusters}",
      journal = {arXiv e-prints},
         year = 2024,
        month = nov,
          eid = {arXiv:2411.16555},
        pages = {arXiv:2411.16555},
          doi = {10.48550/arXiv.2411.16555},
archivePrefix = {arXiv},
       eprint = {2411.16555},
 primaryClass = {astro-ph.GA},
       adsurl = {https://ui.adsabs.harvard.edu/abs/2024arXiv241116555P}
}

@ARTICLE{Ramos-Ceja2025,
       author = {{Ramos-Ceja}, M.~E. and {Fiorino}, L. and {Bulbul}, E. and {Ghirardini}, V. and {Clerc}, N. and {Liu}, A. and {Sanders}, J.~S. and {Bahar}, Y.~E. and {Dietl}, J. and {Kluge}, M. and {Pacaud}, F. and {Artis}, E. and {Balzer}, F. and {Comparat}, J. and {Ding}, Z. and {Malavasi}, N. and {Merloni}, A. and {Mistele}, T. and {Nandra}, K. and {Seppi}, R. and {Zelmer}, S. and {Zhang}, X.},
        title = "{The SRG/eROSITA all-sky survey: X-ray scaling relations of galaxy groups and clusters in the western Galactic hemisphere}",
      journal = {arXiv e-prints},
         year = 2025,
        month = nov,
          eid = {arXiv:2511.14356},
        pages = {arXiv:2511.14356},
          doi = {10.48550/arXiv.2511.14356},
archivePrefix = {arXiv},
       eprint = {2511.14356},
 primaryClass = {astro-ph.CO},
       adsurl = {https://ui.adsabs.harvard.edu/abs/2025arXiv251114356R}
}

@ARTICLE{Rasia2006,
       author = {{Rasia}, E. and {Ettori}, S. and {Moscardini}, L. and {Mazzotta}, P. and {Borgani}, S. and {Dolag}, K. and {Tormen}, G. and {Cheng}, L.~M. and {Diaferio}, A.},
        title = "{Systematics in the X-ray cluster mass estimators}",
      journal = {\mnras},
         year = 2006,
        month = jul,
       volume = {369},
       number = {4},
        pages = {2013-2024},
          doi = {10.1111/j.1365-2966.2006.10466.x},
archivePrefix = {arXiv},
       eprint = {astro-ph/0602434},
 primaryClass = {astro-ph},
       adsurl = {https://ui.adsabs.harvard.edu/abs/2006MNRAS.369.2013R}
}

@ARTICLE{Rephaeli2008,
       author = {{Rephaeli}, Y. and {Nevalainen}, J. and {Ohashi}, T. and {Bykov}, A.~M.},
        title = "{Nonthermal Phenomena in Clusters of Galaxies}",
      journal = {\ssr},
         year = 2008,
        month = feb,
       volume = {134},
       number = {1-4},
        pages = {71-92},
          doi = {10.1007/s11214-008-9314-7},
archivePrefix = {arXiv},
       eprint = {0801.0982},
 primaryClass = {astro-ph},
       adsurl = {https://ui.adsabs.harvard.edu/abs/2008SSRv..134...71R}
}

@ARTICLE{Siegel2025,
       author = {{Siegel}, Jared and {Bigwood}, Leah and {Amon}, Alexandra and {McCullough}, Jamie and {Yamamoto}, Masaya and {McCarthy}, Ian G. and {Schaller}, Matthieu and {Schneider}, Aurel and {Schaye}, Joop},
        title = "{The suppression of the matter power spectrum: strong feedback from X-ray gas mass fractions, kSZ effect profiles, and galaxy-galaxy lensing}",
      journal = {arXiv e-prints},
         year = 2025,
        month = dec,
          eid = {arXiv:2512.02954},
        pages = {arXiv:2512.02954},
          doi = {10.48550/arXiv.2512.02954},
archivePrefix = {arXiv},
       eprint = {2512.02954},
 primaryClass = {astro-ph.CO},
       adsurl = {https://ui.adsabs.harvard.edu/abs/2025arXiv251202954S}
}

@ARTICLE{Quataert2025,
       author = {{Quataert}, Eliot and {Hopkins}, Philip F.},
        title = "{Cosmic Ray Feedback in Massive Halos: Implications for the Distribution of Baryons}",
      journal = {The Open Journal of Astrophysics},
         year = 2025,
        month = may,
       volume = {8},
          eid = {66},
        pages = {66},
          doi = {10.33232/001c.138772},
archivePrefix = {arXiv},
       eprint = {2502.01753},
 primaryClass = {astro-ph.CO},
       adsurl = {https://ui.adsabs.harvard.edu/abs/2025OJAp....8E..66Q}
}

@ARTICLE{Randall2015,
       author = {{Randall}, S.~W. and {Nulsen}, P.~E.~J. and {Jones}, C. and {Forman}, W.~R. and {Bulbul}, E. and {Clarke}, T.~E. and {Kraft}, R. and {Blanton}, E.~L. and {David}, L. and {Werner}, N. and {Sun}, M. and {Donahue}, M. and {Giacintucci}, S. and {Simionescu}, A.},
        title = "{A Very Deep Chandra Observation of the Galaxy Group NGC 5813: AGN Shocks, Feedback, and Outburst History}",
      journal = {\apj},
         year = 2015,
        month = jun,
       volume = {805},
       number = {2},
          eid = {112},
        pages = {112},
          doi = {10.1088/0004-637X/805/2/112},
archivePrefix = {arXiv},
       eprint = {1503.08205},
 primaryClass = {astro-ph.HE},
       adsurl = {https://ui.adsabs.harvard.edu/abs/2015ApJ...805..112R}
}

@article{Randall2018,
       author = {{Randall}, Scott W. and {Nulsen}, Paul E. J. and {Jones}, Christine and {Machacek}, Marie E. and {Blanton}, Elizabeth L. and {O'Sullivan}, Ewan and {Forman}, William R. and {Bulbul}, Esra and {David}, Laurence P.},
        title = "{AGN Feedback and Gas Sloshing in the Nearby Galaxy Group NGC 5044}",
      journal = {\apj},
         year = 2018,
        month = sep,
       volume = {864},
       number = {1},
        pages = {11},
          doi = {10.3847/1538-4357/aad2b9},
       adsurl = {https://ui.adsabs.harvard.edu/abs/2018ApJ...864...11R}
}

@ARTICLE{Reiprich2021,
       author = {{Reiprich}, T.~H. and {Veronica}, A. and {Pacaud}, F. and {Ramos-Ceja}, M.~E. and {Ota}, N. and {Sanders}, J. and {Kara}, M. and {Erben}, T. and {Klein}, M. and {Erler}, J. and {Kerp}, J. and {Hoang}, D.~N. and {Br{\"u}ggen}, M. and {Marvil}, J. and {Rudnick}, L. and {Biffi}, V. and {Dolag}, K. and {Aschersleben}, J. and {Basu}, K. and {Brunner}, H. and {Bulbul}, E. and {Dennerl}, K. and {Eckert}, D. and {Freyberg}, M. and {Gatuzz}, E. and {Ghirardini}, V. and {K{\"a}fer}, F. and {Merloni}, A. and {Migkas}, K. and {Nandra}, K. and {Predehl}, P. and {Robrade}, J. and {Salvato}, M. and {Whelan}, B. and {Diaz-Ocampo}, A. and {Hernandez-Lang}, D. and {Zenteno}, A. and {Brown}, M.~J.~I. and {Collier}, J.~D. and {Diego}, J.~M. and {Hopkins}, A.~M. and {Kapinska}, A. and {Koribalski}, B. and {Mroczkowski}, T. and {Norris}, R.~P. and {O'Brien}, A. and {Vardoulaki}, E.},
        title = "{The Abell 3391/95 galaxy cluster system. A 15 Mpc intergalactic medium emission filament, a warm gas bridge, infalling matter clumps, and (re-) accelerated plasma discovered by combining SRG/eROSITA data with ASKAP/EMU and DECam data}",
      journal = {\aap},
         year = 2021,
        month = mar,
       volume = {647},
          eid = {A2},
        pages = {A2},
          doi = {10.1051/0004-6361/202039590},
archivePrefix = {arXiv},
       eprint = {2012.08491},
 primaryClass = {astro-ph.CO},
       adsurl = {https://ui.adsabs.harvard.edu/abs/2021A&A...647A...2R}
}

@ARTICLE{Ruszkowski2023,
       author = {{Ruszkowski}, Mateusz and {Pfrommer}, Christoph},
        title = "{Cosmic ray feedback in galaxies and galaxy clusters}",
      journal = {\aapr},
         year = 2023,
        month = dec,
       volume = {31},
       number = {1},
          eid = {4},
        pages = {4},
          doi = {10.1007/s00159-023-00149-2},
archivePrefix = {arXiv},
       eprint = {2306.03141},
 primaryClass = {astro-ph.HE},
       adsurl = {https://ui.adsabs.harvard.edu/abs/2023A&ARv..31....4R}
}

@ARTICLE{Sarkar2022,
       author = {{Sarkar}, Arnab and {Su}, Yuanyuan and {Truong}, Nhut and {Randall}, Scott and {Mernier}, Fran{\c{c}}ois and {Gastaldello}, Fabio and {Biffi}, Veronica and {Kraft}, Ralph},
        title = "{Chemical abundances in the outskirts of nearby galaxy groups measured with joint Suzaku and Chandra observations}",
      journal = {\mnras},
         year = 2022,
        month = oct,
       volume = {516},
       number = {2},
        pages = {3068-3081},
          doi = {10.1093/mnras/stac2416},
archivePrefix = {arXiv},
       eprint = {2208.06085},
 primaryClass = {astro-ph.GA},
       adsurl = {https://ui.adsabs.harvard.edu/abs/2022MNRAS.516.3068S}
}

@ARTICLE{Schaye2023,
       author = {{Schaye}, Joop and {Kugel}, Roi and {Schaller}, Matthieu and {Helly}, John C. and {Braspenning}, Joey and {Elbers}, Willem and {McCarthy}, Ian G. and {van Daalen}, Marcel P. and {Vandenbroucke}, Bert and {Frenk}, Carlos S. and {Kwan}, Juliana and {Salcido}, Jaime and {Bah{\'e}}, Yannick M. and {Borrow}, Josh and {Chaikin}, Evgenii and {Hahn}, Oliver and {Hu{\v{s}}ko}, Filip and {Jenkins}, Adrian and {Lacey}, Cedric G. and {Nobels}, Folkert S.~J.},
        title = "{The FLAMINGO project: cosmological hydrodynamical simulations for large-scale structure and galaxy cluster surveys}",
      journal = {\mnras},
         year = 2023,
        month = dec,
       volume = {526},
       number = {4},
        pages = {4978-5020},
          doi = {10.1093/mnras/stad2419},
archivePrefix = {arXiv},
       eprint = {2306.04024},
 primaryClass = {astro-ph.CO},
       adsurl = {https://ui.adsabs.harvard.edu/abs/2023MNRAS.526.4978S}
}

@ARTICLE{Simionescu2019,
       author = {{Simionescu}, A. and {Nakashima}, S. and {Yamaguchi}, H. and {Matsushita}, K. and {Mernier}, F. and {Werner}, N. and {Tamura}, T. and {Nomoto}, K. and {de Plaa}, J. and {Leung}, S.-C. and {Bamba}, A. and {Bulbul}, E. and {Eckart}, M.~E. and {Ezoe}, Y. and {Fabian}, A.~C. and {Fukazawa}, Y. and {Gu}, L. and {Ichinohe}, Y. and {Ishigaki}, M.~N. and {Kaastra}, J.~S. and {Kilbourne}, C. and {Kitayama}, T. and {Leutenegger}, M. and {Loewenstein}, M. and {Maeda}, Y. and {Miller}, E.~D. and {Mushotzky}, R.~F. and {Noda}, H. and {Pinto}, C. and {Porter}, F.~S. and {Safi-Harb}, S. and {Sato}, K. and {Takahashi}, T. and {Ueda}, S. and {Zha}, S.},
        title = "{Constraints on the chemical enrichment history of the Perseus Cluster of galaxies from high-resolution X-ray spectroscopy}",
      journal = {\mnras},
         year = 2019,
        month = feb,
       volume = {483},
       number = {2},
        pages = {1701-1721},
          doi = {10.1093/mnras/sty3220},
archivePrefix = {arXiv},
       eprint = {1806.00932},
 primaryClass = {astro-ph.HE},
       adsurl = {https://ui.adsabs.harvard.edu/abs/2019MNRAS.483.1701S}
}

@ARTICLE{Sijacki2007,
       author = {{Sijacki}, Debora and {Springel}, Volker and {Di Matteo}, Tiziana and {Hernquist}, Lars},
        title = "{A unified model for AGN feedback in cosmological simulations of structure formation}",
      journal = {\mnras},
         year = 2007,
        month = sep,
       volume = {380},
       number = {3},
        pages = {877-900},
          doi = {10.1111/j.1365-2966.2007.12153.x},
archivePrefix = {arXiv},
       eprint = {0705.2238},
 primaryClass = {astro-ph},
       adsurl = {https://ui.adsabs.harvard.edu/abs/2007MNRAS.380..877S}
}

@ARTICLE{Springel2018,
       author = {{Springel}, Volker and {Pakmor}, R{\"u}diger and {Pillepich}, Annalisa and {Weinberger}, Rainer and {Nelson}, Dylan and {Hernquist}, Lars and {Vogelsberger}, Mark and {Genel}, Shy and {Torrey}, Paul and {Marinacci}, Federico and {Naiman}, Jill},
        title = "{First results from the IllustrisTNG simulations: matter and galaxy clustering}",
      journal = {\mnras},
         year = 2018,
        month = mar,
       volume = {475},
       number = {1},
        pages = {676-698},
          doi = {10.1093/mnras/stx3304},
archivePrefix = {arXiv},
       eprint = {1707.03397},
 primaryClass = {astro-ph.GA},
       adsurl = {https://ui.adsabs.harvard.edu/abs/2018MNRAS.475..676S}
}

@ARTICLE{Sun2009,
       author = {{Sun}, M. and {Voit}, G.~M. and {Donahue}, M. and {Jones}, C. and {Forman}, W. and {Vikhlinin}, A.},
        title = "{Chandra Studies of the X-Ray Gas Properties of Galaxy Groups}",
      journal = {\apj},
         year = 2009,
        month = mar,
       volume = {693},
       number = {2},
        pages = {1142-1172},
          doi = {10.1088/0004-637X/693/2/1142},
archivePrefix = {arXiv},
       eprint = {0805.2320},
 primaryClass = {astro-ph},
       adsurl = {https://ui.adsabs.harvard.edu/abs/2009ApJ...693.1142S}
}

@ARTICLE{Sunyaev2021,
       author = {{Sunyaev}, R. and {Arefiev}, V. and {Babyshkin}, V. and {Bogomolov}, A. and {Borisov}, K. and {Buntov}, M. and {Brunner}, H. and {Burenin}, R. and {Churazov}, E. and {Coutinho}, D. and {Eder}, J. and {Eismont}, N. and {Freyberg}, M. and {Gilfanov}, M. and {Gureyev}, P. and {Hasinger}, G. and {Khabibullin}, I. and {Kolmykov}, V. and {Komovkin}, S. and {Krivonos}, R. and {Lapshov}, I. and {Levin}, V. and {Lomakin}, I. and {Lutovinov}, A. and {Medvedev}, P. and {Merloni}, A. and {Mernik}, T. and {Mikhailov}, E. and {Molodtsov}, V. and {Mzhelsky}, P. and {M{\"u}ller}, S. and {Nandra}, K. and {Nazarov}, V. and {Pavlinsky}, M. and {Poghodin}, A. and {Predehl}, P. and {Robrade}, J. and {Sazonov}, S. and {Scheuerle}, H. and {Shirshakov}, A. and {Tkachenko}, A. and {Voron}, V.},
        title = "{SRG X-ray orbital observatory. Its telescopes and first scientific results}",
      journal = {\aap},
         year = 2021,
        month = dec,
       volume = {656},
          eid = {A132},
        pages = {A132},
          doi = {10.1051/0004-6361/202141179},
archivePrefix = {arXiv},
       eprint = {2104.13267},
 primaryClass = {astro-ph.HE},
       adsurl = {https://ui.adsabs.harvard.edu/abs/2021A&A...656A.132S}
}

@ARTICLE{Urban2017,
       author = {{Urban}, O. and {Werner}, N. and {Allen}, S.~W. and {Simionescu}, A. and {Mantz}, A.},
        title = "{A uniform metallicity in the outskirts of massive, nearby galaxy clusters}",
      journal = {\mnras},
         year = 2017,
        month = oct,
       volume = {470},
       number = {4},
        pages = {4583-4599},
          doi = {10.1093/mnras/stx1542},
archivePrefix = {arXiv},
       eprint = {1706.01567},
 primaryClass = {astro-ph.CO},
       adsurl = {https://ui.adsabs.harvard.edu/abs/2017MNRAS.470.4583U}
}

@ARTICLE{Vernstrom2021,
       author = {{Vernstrom}, T. and {Heald}, G. and {Vazza}, F. and {Galvin}, T.~J. and {West}, J.~L. and {Locatelli}, N. and {Fornengo}, N. and {Pinetti}, E.},
        title = "{Discovery of magnetic fields along stacked cosmic filaments as revealed by radio and X-ray emission}",
      journal = {\mnras},
         year = 2021,
        month = aug,
       volume = {505},
       number = {3},
        pages = {4178-4196},
          doi = {10.1093/mnras/stab1301},
archivePrefix = {arXiv},
       eprint = {2101.09331},
 primaryClass = {astro-ph.CO},
       adsurl = {https://ui.adsabs.harvard.edu/abs/2021MNRAS.505.4178V}
}

@ARTICLE{Vernstrom2023,
       author = {{Vernstrom}, Tessa and {West}, Jennifer and {Vazza}, Franco and {Wittor}, Denis and {Riseley}, Christopher John and {Heald}, George},
        title = "{Polarized accretion shocks from the cosmic web}",
      journal = {Science Advances},
         year = 2023,
        month = feb,
       volume = {9},
       number = {7},
          eid = {eade7233},
        pages = {eade7233},
          doi = {10.1126/sciadv.ade7233},
archivePrefix = {arXiv},
       eprint = {2302.08072},
 primaryClass = {astro-ph.CO},
       adsurl = {https://ui.adsabs.harvard.edu/abs/2023SciA....9E7233V}
}

@ARTICLE{Veronica2024,
       author = {{Veronica}, Angie and {Reiprich}, Thomas H. and {Pacaud}, Florian and {Ota}, Naomi and {Aschersleben}, Jann and {Biffi}, Veronica and {Bulbul}, Esra and {Clerc}, Nicolas and {Dolag}, Klaus and {Erben}, Thomas and {Gatuzz}, Efrain and {Ghirardini}, Vittorio and {Kerp}, J{\"u}rgen and {Klein}, Matthias and {Liu}, Ang and {Liu}, Teng and {Migkas}, Konstantinos and {Ramos-Ceja}, Miriam E. and {Sanders}, Jeremy and {Spinelli}, Claudia},
        title = "{The eROSITA view of the Abell 3391/95 field. Cluster outskirts and filaments}",
      journal = {\aap},
         year = 2024,
        month = jan,
       volume = {681},
          eid = {A108},
        pages = {A108},
          doi = {10.1051/0004-6361/202347037},
archivePrefix = {arXiv},
       eprint = {2311.07488},
 primaryClass = {astro-ph.CO},
       adsurl = {https://ui.adsabs.harvard.edu/abs/2024A&A...681A.108V}
}

@ARTICLE{Wik2014,
       author = {{Wik}, Daniel R. and {Hornstrup}, A. and {Molendi}, S. and {Madejski}, G. and {Harrison}, F.~A. and {Zoglauer}, A. and {Grefenstette}, B.~W. and {Gastaldello}, F. and {Madsen}, K.~K. and {Westergaard}, N.~J. and {Ferreira}, D.~D.~M. and {Kitaguchi}, T. and {Pedersen}, K. and {Boggs}, S.~E. and {Christensen}, F.~E. and {Craig}, W.~W. and {Hailey}, C.~J. and {Stern}, D. and {Zhang}, W.~W.},
        title = "{NuSTAR Observations of the Bullet Cluster: Constraints on Inverse Compton Emission}",
      journal = {\apj},
         year = 2014,
        month = sep,
       volume = {792},
       number = {1},
          eid = {48},
        pages = {48},
          doi = {10.1088/0004-637X/792/1/48},
archivePrefix = {arXiv},
       eprint = {1403.2722},
 primaryClass = {astro-ph.HE},
       adsurl = {https://ui.adsabs.harvard.edu/abs/2014ApJ...792...48W}
}

@ARTICLE{Wik2012,
       author = {{Wik}, Daniel R. and {Sarazin}, Craig L. and {Zhang}, Yu-Ying and {Baumgartner}, Wayne H. and {Mushotzky}, Richard F. and {Tueller}, Jack and {Okajima}, Takashi and {Clarke}, Tracy E.},
        title = "{The Swift Burst Alert Telescope Perspective on Non-thermal Emission in HIFLUGCS Galaxy Clusters}",
      journal = {\apj},
         year = 2012,
        month = mar,
       volume = {748},
       number = {1},
          eid = {67},
        pages = {67},
          doi = {10.1088/0004-637X/748/1/67},
archivePrefix = {arXiv},
       eprint = {1207.0506},
 primaryClass = {astro-ph.CO},
       adsurl = {https://ui.adsabs.harvard.edu/abs/2012ApJ...748...67W}
}

@INPROCEEDINGS{Zabalza2015,
       author = {{Zabalza}, V.},
        title = "{Naima: a Python package for inference of particle distribution properties from nonthermal spectra}",
    booktitle = {34th International Cosmic Ray Conference (ICRC2015)},
         year = 2015,
       series = {International Cosmic Ray Conference},
       volume = {34},
        month = jul,
          eid = {922},
        pages = {922},
          doi = {10.22323/1.236.0922},
archivePrefix = {arXiv},
       eprint = {1509.03319},
 primaryClass = {astro-ph.HE},
       adsurl = {https://ui.adsabs.harvard.edu/abs/2015ICRC...34..922Z}
}

@ARTICLE{Zhang2024,
       author = {{Zhang}, X. and {Bulbul}, E. and {Malavasi}, N. and {Ghirardini}, V. and {Comparat}, J. and {Kluge}, M. and {Liu}, A. and {Merloni}, A. and {Zhang}, Y. and {Bahar}, Y.~E. and {Artis}, E. and {Sanders}, J.~S. and {Garrel}, C. and {Balzer}, F. and {Br{\"u}ggen}, M. and {Freyberg}, M. and {Gatuzz}, E. and {Grandis}, S. and {Krippendorf}, S. and {Nandra}, K. and {Ponti}, G. and {Ramos-Ceja}, M. and {Predehl}, P. and {Reiprich}, T.~H. and {Veronica}, A. and {Yeung}, M.~C.~H. and {Zelmer}, S.},
        title = "{The SRG/eROSITA all-sky survey: X-ray emission from the warm-hot phase gas in long cosmic filaments}",
      journal = {\aap},
         year = 2024,
        month = nov,
       volume = {691},
          eid = {A234},
        pages = {A234},
          doi = {10.1051/0004-6361/202450933},
archivePrefix = {arXiv},
       eprint = {2406.00105},
 primaryClass = {astro-ph.HE},
       adsurl = {https://ui.adsabs.harvard.edu/abs/2024A&A...691A.234Z}
}

@ARTICLE{Zhang2025,
       author = {{Zhang}, X. and {Bulbul}, E. and {Diemer}, B. and {Bahar}, Y.~E. and {Comparat}, J. and {Ghirardini}, V. and {Liu}, A. and {Malavasi}, N. and {Mistele}, T. and {Ramos-Ceja}, M. and {Sanders}, J.~S. and {Zhang}, Y. and {Artis}, E. and {Ding}, Z. and {Fiorino}, L. and {Kluge}, M. and {Merloni}, A. and {Nandra}, K. and {Zelmer}, S.},
        title = "{The SRG/eROSITA All-Sky Survey. Detection of shock-heated gas beyond the halo boundary into the accretion region}",
      journal = {arXiv e-prints},
         year = 2025,
        month = sep,
          eid = {arXiv:2509.25317},
        pages = {arXiv:2509.25317},
          doi = {10.48550/arXiv.2509.25317},
archivePrefix = {arXiv},
       eprint = {2509.25317},
 primaryClass = {astro-ph.CO},
       adsurl = {https://ui.adsabs.harvard.edu/abs/2025arXiv250925317Z}
}

\end{document}